\documentclass[%
 reprint,
 prd,
 showpacs,
 amsmath,amssymb,
 aps,
 lengthcheck,
floatfix,
]{revtex4-2}
\usepackage{bibunits}

\usepackage{graphicx}
\usepackage{dcolumn}
\usepackage{bm}
\usepackage{hyperref}
\usepackage[mathlines]{lineno}
\usepackage{lipsum}

\usepackage{soul}

\usepackage{graphicx}
\usepackage{times}
\usepackage[normalem]{ulem}
\usepackage{color}
\usepackage{cellspace}
\usepackage{multirow}

\newcommand{\comment}[1]{}
\renewcommand\sout{\bgroup \color{red} \ULdepth=-.5ex \ULset}

\usepackage{rotating}
\usepackage{adjustbox}
\usepackage{makecell}
\usepackage{xcolor}
\usepackage{amsmath}
\usepackage{orcidlink}
\def\simge{\mathrel{\rlap{\raise 0.511ex
     \hbox{$>$}}{\lower 0.511ex \hbox{$\sim$}}}}
\def\simle{\mathrel{\rlap{\raise 0.511ex
      \hbox{$<$}}{\lower 0.511ex \hbox{$\sim$}}}}

\usepackage{braket}
\newcommand{\su}{\affiliation{ Department of Physics, Syracuse University, Syracuse, NY 13244, USA}}

\newcommand{\princeton}{\affiliation{Department of Astrophysical Sciences, Princeton University, Princeton, NJ 08544}}

\newcommand{\carnegie}{\affiliation{Carnegie Observatories, 813 Santa Barbara St, Pasadena, CA 91101, USA; NASA Hubble Fellow}}

\begin{document}

\title{Measuring the properties of $f-$mode oscillations of a protoneutron star by third generation gravitational-wave detectors}

\author{Chaitanya~Afle$^{1}$\,\orcidlink{0000-0003-2900-1333}}\email{chafle@syr.edu}\su
\author{Suman~Kumar~Kundu\,\orcidlink{0000-0001-6118-0023}}\su
\author{Jenna~Cammerino}\su
\author{Eric~R~Coughlin\,\orcidlink{0000-0003-3765-6401}}\su
\author{Duncan~A.~Brown$^{1}$\, \orcidlink{0000-0002-9180-5765}}\su

\author{David Vartanyan\,\orcidlink{0000-0003-1938-9282}}\carnegie

\author{Adam Burrows\,\orcidlink{0000-0002-3099-5024}}\princeton

\begin{abstract}
Core-collapse supernovae are among the astrophysical sources of gravitational waves that could be detected by third-generation gravitational-wave detectors. Here, we analyze the gravitational-wave strain signals from two- and three-dimensional simulations of core-collapse supernovae generated using the code F{\sc{ornax}}. A subset of the two-dimensional simulations has non-zero core rotation at the core bounce. A dominant source of time changing quadrupole moment is the $l=2$ fundamental mode ($f-$ mode) oscillation of the proto-neutron star. From the time-frequency spectrogram of the gravitational-wave strain we see that, starting $\sim 400$ ms after the core bounce, most of the power lies within a narrow track that represents the frequency evolution of the $f-$mode oscillations. The $f-$mode frequencies obtained from linear perturbation analysis of the angle-averaged profile of the protoneutron star corroborate what we observe in the spectrograms of the gravitational-wave signal. We explore the measurability of the $f-$mode frequency evolution of protoneutron star for a supernova signal observed in the third-generation gravitational-wave detectors. Measurement of the frequency evolution can reveal information about the masses, radii, and densities of the proto-neutron stars. We find that if the third generation detectors observe a supernova within 10 kpc, we can measure these frequencies to within $\sim$90\% accuracy. We can also measure the energy emitted in the fundamental $f-$mode using the spectrogram data of the strain signal. We find that the energy in the $f-$mode can be measured to within 20\% error for signals observed by Cosmic Explorer using simulations with successful explosion, assuming source distances within 10 kpc.
\end{abstract}

\maketitle

\section{Introduction}

The core of a massive star ($M \geq 8 M_{\odot}$) collapses due to gravity upon achieving the effective Chandrasekhar mass of a massive-star progenitor \citep{Bethe:1990mw,Janka:2012wk,Burrows:2012ew}. 
The inner part of the core collapses to nuclear densities to form a proto-neutron star. 
A shockwave is created at the boundary of the protoneutron star and propagates outwards.
The shock is initially stopped in its progress outward as a fraction of the kinetic energy of the shock is used to dissociate the heavy nuclei.
A fraction of neutrinos produced in the proto-neutron star are trapped behind the shock. This heats up the the shocked region and enhances the turbulent convection, which revives the stalled shock \cite{Colgate:1966ax}. 
The joint observation of the photons, neutrinos, and gravitational waves emitted during this process can help reveal the mechanism by which the shock is revived and a neutron star is born. 

Core-collapse supernovae are possible sources of gravitational-waves that could be detected by the proposed third-generation interferometric detectors, such as the Cosmic Explorer \cite{Reitze:2019iox, Evans:2021gyd} and the Einstein Telescope. These observatories will be able to detect a supernova within 100 kpc, which includes the Milky Way galaxy and its satellites \cite{Srivastava:2019fcb}. The estimated rate of supernovae for a galaxy the size of Milky Way is 1-3 per century \cite{adams2013observing,Cappellaro:1996cc,1994ApJS...92..487T}.  

A number of studies have characterized the gravitational wave signal from the collapse and explosion of the core of a massive star \citep{Mueller:97,Kotake:03, Murphy:2009dx,Ott:2012kr,Hayama:15,Yakunin:15,Powell:2016wke,Andresen:17,Morozova:2018glm,Radice2019}. After decades of improvement in the numerical techniques, we are now much better able to account for the complex hydrodynamics in multi-dimensions, the neutrino interactions, and the hydrodynamical instabilities \citep{Burrows:18,Burrows:2019zce,Skinner:19,Burrows:2020qrp, Fryer:2011zz}. 

This rich and complex physics gives rise to a complex gravitational-wave signal, which in the time domain represents the stochastic nature of matter movements within the star. There is a sharp negative peak in the signal at the time of core bounce, and its amplitude depends on the rotation rate of the core of the progenitor. This is followed by the post-bounce oscillations of the core, that extend for ~6-10 ms after the bounce, with an amplitude that depends interestingly on the rotation rate of the core of the progenitor star and its equation of state \citep{Richers:2017joj}. The end of post-bounce oscillations mark the onset of ``prompt convection" due to the dynamical imposition of a negative entropy gradient as the shock stalls. Starting from $\sim$150 ms after the core bounce, there is a strong, stochastic signal. Moreover, an asymmetrical explosion is accompanied by a growing offset in the mean strain from zero due to ``memory"\cite{Christodoulou:1991cr, Thorne:1992sdb, Vartanyan:2020nmt}.     

Even though the signal is highly stochastic in the time domain, the time-frequency spectrogram of the gravitational-wave signal reveals that most of the power lies in a narrow track in the time-frequency plane.  Linear perturbation analysis of the proto-neutron star shows that this frequency corresponds to the quadrupolar $f-$mode of the proto-neutron star \citep{Morozova:2018glm,Radice2019}, which start approximately $100-400$ ms after the core bounce. These oscillations are excited by the downflows of matter accreted onto the proto-neutron star \citep{Murphy:2009dx,Morozova:2018glm}. 

Previously, Ref. \cite{Bizouard:2020sws}  measured the frequencies associated with the $g$-mode oscillations of the proto-neutron star using the time-frequency spectrograms of the gravitational-wave strains obtained from simulations. Using the frequency measurement and universal relations, they obtain measurement of the ratio $M_{\mathrm{PNS}}/R_{\mathrm{PNS}}^2$ of the proto-neutron star, where $M_{\mathrm{PNS}}$ is the mass and $R_{\mathrm{PNS}}$ is the radius of the proto-neutron star. Ref. \cite{Powell:2022nrs} develop a phenomenological model of the gravitational-wave signal associated with the dominant mode and use the spectrogram of the strain to measure f-mode frequency evolution and energy. They use Bayesian parameter estimation to measure their model parameters and then obtain frequencies and energies associated with the mode from the posteriors. More recently, Ref. \cite{Bruel:2023iye} extended the work of Ref. \cite{Bizouard:2020sws} of measuring the $f/g$ mode frequencies of the proto-neutron stars using the strains of 3D and 2D simulations. They use a network of current, and future, detectors to perform a coherent analysis of the detected signal.

In this paper, we develop a model-independent method to measure the $f-$mode frequencies and the energy emitted in gravitational radiation of the proto-neutron star oscillations by analysing the spectrograms of the gravitational-wave strains obtained from state-of-the art three-dimensional core-collapse supernovae simulations. We develop a novel method of generating time-frequency spectrograms that can be used to reliably measure power in a given track on the spectrogram. We inject the strain obtained through the simulations into several instances of simulated detector noise to measure the frequencies and energies. We vary the distance of the source to test this method for signals with various signal-to-noise ratios. We find that, from simulated observations using the third-generation gravitational-wave detectors, while we can detect the signal out to distances of $\approx 100$ kpc, we can measure the frequencies and the energies associated with the $f-$mode oscillations to within $20\%$ error from sources within $\approx 10$ kpc distance.

Section \ref{sec:simulations} describes the numerical simulations used in this paper, and describes the linear perturbation analysis used to determine the $f-$mode oscillation frequencies. In Section \ref{sec:spectrogram_analysis} we describe our method to construct the short-time Fourier transform and the spectrogram of the gravitational-wave signal obtained from the simulations. In Section \ref{sec:results}, we describe our main results from the analysis. We summarize our findings in Section \ref{sec:conclusion}.

\section{Simulations}\label{sec:simulations}

In our analysis we used the data obtained from two- and three-dimensional core-collapse supernovae simulations performed with the neutrino-radiation hydrodynamics code \textsc{FORNAX} \cite{Skinner16,Burrows18,Vartanyan18}. The progenitors used in the simulations were calculated by Refs. \cite{Sukhbold:2015wba} and \cite{Sukhbold:2017cnt}. Further details of the simulations can be found in Refs. \cite{Morozova:2018glm,Burrows:2019zce, Radice2019}. 

We tabulate the models we consider in our works in Table~\ref{tab:simulations}. We show the mass of the progenitor, the equation of state of the proto-neutron star used in the simulations, and the core-rotation rate in columns 3, 4, and 5 of the table. We also indicate whether the the shock is revived and the star explodes within the time of the simulation. For the three-dimensional simulation models, We use a wide range of progenitors with the ZAMS (Zero-Age Main Sequence) mass ranging from $9 M_{\odot} - 60 M_{\odot}$. We use SFHo equation of state, and all but the $13 M_{\odot}, 14 M_{\odot}$, and $15 M_{\odot}$ explode within the time of the simulation. For the two-dimensional simulations with core rotation at the time of core bounce, we use a $15 M_{\odot}$ progenitor. We have a total of 14 models with rotation rates ranging from $0.0$ rad/sec $- 6.14$ rad/sec. We also include 9 two-dimensional simulations with zero core rotation.  

The last three columns show the optimal distances for every simulation, for Advanced LIGO \cite{LIGOScientific:2014pky}, Einstein Telescope \cite{Punturo:2010zz}, and Cosmic Explorer\cite{Evans:2021gyd}. Optimal distance of a source, for a given detector is defined as the distance at which the signal-to-noise ratio of the optimally-oriented source is equal to eight. It is calculated as,
\begin{equation}
    d_{\mathrm{opt}} = \frac{1}{\rho_{\mathrm{opt}}}\left[2 \int_{f_{\mathrm{low}}}^{f_{\mathrm{high}}}df \frac{\Tilde{h}(f)\Tilde{h}^{*}(f)}{S_n(f)}\right],
\end{equation}
where $\rho_{\mathrm{opt}}= 8$ is the signal-to-noise ratio of an optimal detection, $\Tilde{h}(f)$ is the strain signal in the Fourier domain, and $S_n (f)$ is the power spectral density of the detector noise. For Advanced LIGO, we use the \texttt{aLIGOZeroDetHighPower} \cite{lalsuite} power spectral density, with $f_{\mathrm{low}} = 10$ Hz. The average of the optimal distances of the waveforms from three-dimensional simulations for Advanced LIGO is 8 kpc. Hence, we can detect a signal coming from the center of the galaxy if it is loud enough. The next generation detectors, Einstein Telescope and Cosmic Explorer, can detect signals coming from the Milky Way galaxy and its satellite galaxies. Their detection range is large enough to cover the entire Milky way but not large enough to reach the nearest galaxy, Andromeda, which is at 770 kpc. The gravitational-wave signals from core-collapse supernovae observed by the third-generation detectors will have large signal-to-noise ratio.    

\begin{table*}
\centering
\begin{tabular}{|c|c|c|c|c|c|c|c|c|}
\hline
\hline
\multirow{2}{*}{}&\multirow{2}{*}{Label} & \multirow{2}{*}{\shortstack[c]{\\Progenitor \\ mass \\($M_{\odot}$)}} & \multirow{2}{*}{\shortstack[c]{Equation \\ of state}} &
\multirow{2}{*}{\shortstack[c]{\\Core \\ rotation \\ rate (rad/sec) }} & \multirow{2}{*}{\shortstack[c]{Explosion\\ Status}} &
 \multicolumn{3}{c|}{Optimal Distance (kpc)} \\
\cline{7-9}
 & &  & &  &  & \makecell{Advanced \\LIGO} & \shortstack[c]{Einstein Telescope}  & \shortstack[c]{Cosmic Explorer} \\
\hline

\multirow{6}{*}{\shortstack[c]{Three-dimensional \\ simulations}}
                           & \texttt{s9-3D}   & 9  & SFHo & - & Yes & 2  & 25  & 39  \\
                           & \texttt{s10-3D}  & 10 & SFHo & - & Yes & 7  & 73  & 116 \\
                           & \texttt{s11-3D}  & 11 & SFHo & - & Yes & 6  & 61  & 98  \\
                           & \texttt{s12-3D}  & 12 & SFHo & - & Yes & 8  & 79  & 127 \\
                           & \texttt{s13-3D}  & 13 & SFHo & - & No  & 7  & 71  & 118 \\                & \texttt{s14-3D}  & 14 & SFHo & - & No  & 7  & 74  & 121 \\
                           & \texttt{s15-3D}  & 15 & SFHo & - & No  & 7  & 72  & 114 \\
                           & \texttt{s17-3D}  & 17 & SFHo & - & Yes  & 11 & 107 & 171 \\
                           & \texttt{s18-3D}  & 18 & SFHo & - & Yes  & 11 & 108 & 174 \\
                           & \texttt{s19-3D}  & 19 & SFHo & - & Yes & 15 & 141 & 228 \\
                           & \texttt{s20-3D}  & 20 & SFHo & - & Yes  & 13 & 131 & 214 \\
                           & \texttt{s25-3D}  & 25 & SFHo & - & Yes & 13 & 125 & 208 \\
                           & \texttt{s60-3D}  & 60 & SFHo & - & Yes & 9  & 93  & 150 \\
\hline

\multirow{6}{*}{\shortstack[c]{Two-dimensional \\ simulations \\ with \\ core rotation}} 
                & \texttt{0.0strain}  & 15 & SFHo & 0.0  & No   & 28  & 270  & 427   \\
                & \texttt{0.05strain} & 15 & SFHo & 0.05 & No   & 32  & 324  & 516   \\
                & \texttt{0.1strain}  & 15 & SFHo & 0.1  & No   & 32  & 320  & 531   \\
                & \texttt{0.2strain}  & 15 & SFHo & 0.2  & No   & 45  & 458  & 735   \\     
                & \texttt{0.25strain} & 15 & SFHo & 0.25 & No   & 46  & 466  & 797   \\
                & \texttt{0.3strain}  & 15 & SFHo & 0.3  & No   & 34  & 335  & 572   \\ 
                & \texttt{0.4strain}  & 15 & SFHo & 0.4  & No   & 51  & 510  & 863   \\
                & \texttt{0.5strain}  & 15 & SFHo & 0.5  & No   & 54  & 534  & 903   \\
                & \texttt{0.75strain} & 15 & SFHo & 0.75 & No   & 59  & 567  & 907   \\
                & \texttt{1.0strain}  & 15 & SFHo & 1.0  & No   & 79  & 763  & 1249  \\
                & \texttt{2.0strain}  & 15 & SFHo & 2.0  & No   & 106  & 1096 & 1715  \\
                & \texttt{pi.strain}  & 15 & SFHo & 3.14 & Yes  & 140 & 1465 & 2343  \\
                & \texttt{4.0strain}  & 15 & SFHo & 4.0  & No   & 145  & 1539 & 2511  \\
                & \texttt{5.0strain}  & 15 & SFHo & 5.0  & No   & 146  & 1580 & 2633  \\
                & \texttt{2pi.strain} & 15 & SFHo & 6.28 & No   &  123 & 1312 & 2316  \\
\hline

\multirow{6}{*}{\shortstack[c]{Two-dimensional \\ simulations \\ without \\ core rotation}} 
                            & \texttt{M10-LS220}  & 10 & LS220 & - & No  & 15 & 150 & 232 \\
                            & \texttt{M10-DD2}    & 10 & DD2   & - & No  & 17 & 176 & 268 \\
                            & \texttt{M10-SFHo}   & 10 & SFHo  & - & Yes & 36 & 361 & 566 \\
                            & \texttt{M13-SFHo}   & 13 & SFHo  & - & No  & 30 & 312 & 483 \\
                            & \texttt{M19-SFHo}   & 19 & SFHo  & - & Yes & 55 & 534 & 880 \\
                            & \texttt{gw-s11-2D}  & 11 & SFHo  & - & No  & 31 & 307 & 481 \\
                            & \texttt{gw-s19-2D}  & 19 & SFHo  & - & No  & 39 & 401 & 621 \\
                            & \texttt{gw-s25-2D}  & 25 & SFHo  & - & No  & 47 & 446 & 715 \\
                            & \texttt{gw-s60-2D}  & 60 & SFHo  & - & No  & 47 & 460 & 732 \\

\hline
\hline
\end{tabular}
\caption{The table summarizes the details of the simulations, including the progenitor mass, equation of state, initial core rotation, and explosion status within the simulated time interval. Based on the gravitational-wave strain obtained from the simulations, we also measure the optimal distance of the strain signal for Advanced LIGO, Einstein Telescope, Cosmic Explorer.} \label{tab:simulations}
\end{table*}

\subsection{Linear Perturbation Analysis}
In this section, we outline the method of the linear perturbation analysis of the angle-averaged  data of the proto-neutron star profile (i.e. integrated over the solid angle $\Omega$). The proto-neutron star is to be modeled with the energy-momentum tensor of a perfect fluid.
\begin{equation}
    T_{\mu\nu}= \rho H u^{\mu} u^{\nu} + P g_{\mu \nu} ,
\end{equation}
where $\rho$ denotes the rest-mass density, $P$ the pressure, $u^{\mu}$ the fluid $4$-velocity and $H:=(1+\epsilon+P/\rho)$ the specific enthalpy, $\epsilon$ the specific internal energy.  
Under the assumption of spherical symmetry \footnote{The asymmetries are small enough and hence the angle-averaged background can well be approximated as spherically symmetric.}, the space-time metric $g_{\mu \nu}$ in isotropic coordinates , using the (3 + 1) foliation, can be written as \citep[e.g]{Banyuls:97,Tamborra:2017ull,Burrows:18}  ,
\begin{equation}
    ds^2=g_{\mu\nu} dx^{\mu} dx^{\nu}=-\alpha^2 dt^2+ \psi^4 f_{ij} dx^i dx^j
\end{equation}
where $\alpha$ is the lapse function and the metric for spatial slices is approximated to be conformally related to the flat metric $\delta_{ij}$ with a conformal factor $\psi^4$, set to $1$ in all simulations of Table~\ref{tab:simulations}. 

We now perform perturbation analysis on top of this conformally-flat background by linearizing the equations of general relativistic hydrodynamics. In general, the three components of the Lagrangian fluid displacement field, $\boldsymbol{\xi}(\boldsymbol{r},t) \equiv \xi^r \hat{r} + \xi^\theta \hat{\theta}+ \xi^\phi \hat{\phi}$ representing the perturbation, can be resolved in terms of three scalar functions by virtue of the Helmholtz decomposition theorem. If one assumes that the radial component of the fluid vorticity equation vanishes at all points of the star, that is, $(\nabla \times \xi)_r=0$, then one can show that the three components of $\xi$ can now instead be resolved in terms of only two scalar functions. We now decompose these two scalar functions into purely radial functions $(\eta_r(r),\eta_\perp(r))$ supplemented with the spherical harmonics $Y_{lm}$ and mode frequency $\sigma$ as 
\begin{align}
\begin{split}
    \xi^r &= \eta_r Y_{lm} e^{-i\sigma t}, \\
    \xi^\theta &= \eta_\perp \frac{1}{r^2} \partial_\theta Y_{lm} e^{-i\sigma t}, \\
    \xi^\phi &= \eta_\perp \frac{1}{(r\sin{\theta})^2} \partial_\phi Y_{lm} e^{-i\sigma t} .
    \end{split}
\end{align}
Here, any time dependence of the background state is assumed to be very small compared to the eigen value (i.e. the time derivative of any quantity f, $\partial  f/ \partial t << f/\sigma$). If $\sigma$ is real, the system is neutrally stable (i.e. the modes are oscillatory in nature).  As the background metric is assumed to be conformally flat, the perturbation of the metric is accomplished by perturbing the lapse function. Decomposing the perturbation to the lapse function in purely radial and spherical harmonics yields
\begin{equation}
    \delta \alpha= \delta \hat{\alpha} (r) Y_{lm} e^{-i\sigma t} .
\end{equation}
We define $f_\alpha \equiv \partial_r (\hat{\delta \alpha}/\alpha)$; together with $\hat{\delta \alpha}$ it represents the perturbation in the gravity sector. The timescale associated with neutrino heating and nuclear dissociation is typically $>> 1/\sigma$, hence the 
 perturbations to the fluid properties to be adiabatic in nature, implying,
\begin{equation}
    \frac{\partial P}{\partial \rho}|_{\mathrm{adiabatic}}=Hc_s^2 = \frac{P}{\rho} \Gamma_1,
\end{equation}
$c_s$ the relativistic sound speed in the fluid, and $\Gamma_1$ the adiabatic index.
Now the equations of general-relativistic hydrodynamics together with the $00$ component of the Einstein equation can be linearized to obtain the following system of equations:

\begin{multline}
     \partial_r \eta_r + \left[\frac{2}{r} + \frac{1}{\Gamma_1} \frac{\partial_r P }{P} + 6 \frac{\partial_r \psi}{\psi}\right] \eta_r + \\ \frac{\psi^4}{\alpha^2 c_s^2} \left(\sigma^2 - \mathcal{L}^2\right) \eta_{\perp} - \frac{1}{\alpha c_s^2} \delta \hat{\alpha} = 0,
     \label{eq:p1}
\end{multline}

\begin{multline}
    \partial_r \eta_{\perp} - \left(1 - \frac{\mathcal{N}^2}{\sigma^2}\right)\eta_r + \left[\partial_r \ln q - \tilde{G}\left(1 + \frac{1}{c^2_s}\right)\right]\eta_{\perp} - \\ \frac{1}{\alpha \tilde{G}}\frac{\mathcal{N}^2}{\sigma^2} \delta \hat{\alpha} = 0,
\end{multline}

\begin{multline}
    \partial_r f_{\alpha} + 4 \pi \left[ \partial_r \rho - \frac{\rho}{P \Gamma_1} \partial_r P \right] \eta_r - \frac{4 \pi \rho}{P \Gamma_1} q \sigma^2 \eta_{\perp} + \\ \left[ \frac{4\pi\rho^2h}{P \Gamma_1 \alpha} - \frac{1}{\alpha}\frac{l(l+1)}{r^2}\right]\delta \hat{\alpha} = 0,
\end{multline}
and
\begin{equation}
    \partial_r \delta\hat{\alpha} = f_{\alpha} \alpha - \tilde{G}\delta \hat{\alpha} \label{eq:p4} .
\end{equation}

In Equations \ref{eq:p1}-\ref{eq:p4}, we have collected the combination $\rho h \alpha^{-2} \psi^4$ as $q$, $\Tilde{G}$ is the radial component of gravitational acceleration $\Tilde{G}:=-\partial_r \ln{\alpha}$, $\mathcal{N}$ is the relativistic Brunt-V{\"{a}}is{\"{a}}l{\"{a}} frequency,
\begin{equation}
    \mathcal{N}^2=\frac{\alpha \delta_r \alpha}{\psi^4} \left(\frac{1}{\Gamma_1}\frac{\partial_r P}{P}-\frac{\partial_r e}{\rho H}\right)
\end{equation}
and $\mathcal{L}$ is the relativistic Lamb shift,
\begin{equation}
    \mathcal{L}^2=\frac{\alpha^2}{\psi^4}c_s^2\frac{l(l+1)}{r^2}
\end{equation}
The system of equations \ref{eq:p1}-\ref{eq:p4} can be solved by incorporating appropriate boundary conditions: at the outer boundary, set at the radial coordinate where the density $\rho=10^{10} g/cm^{-3}$ we consider the Lagrangian pressure to vanish and at the inner boundary (i.e. $r=0$ ) use the regularity condition  of \citet{Reis92}. Mathematically, this reads, at the outer boundary,
\begin{equation}
    q \sigma^2 \eta_{\perp} - \frac{\rho H}{\alpha} \hat{\delta \alpha} +\partial_r P \eta_r=0
    \label{eq:obc}
\end{equation}
and at the inner boundary,
\begin{align*}
    \eta_r=\frac{l}{r} \eta_{\perp} \propto r^{l-1} \\
    \eta_r|_{r=0}=\eta_{\perp}|_{r=0}=0
\end{align*}
By discretizing the derivatives by means of trapezoidal rules, we can start integrating the set of equations \ref{eq:p1}-\ref{eq:p4} by inverting the $4 \times 4$ coefficient matrix at every step to solve for $(\eta_r,\eta_{\perp},f_{\alpha},\hat{\delta \alpha})$ and then using the bisection method to uniquely determine the solutions by satisfying the outer boundary condition Eq. \ref{eq:obc}. The eigenvalue corresponding to the unique solution thus obtained gives the frequency of oscillation as $\sigma/2\pi$. The lowest frequency oscillation mode is the fundamental oscillation mode (f-mode), with zero radial nodes. We find the f-mode starts few hundred $ms$ after the core bounce for the simulations in Table~\ref{tab:simulations} which confirms similar findings in \cite{Morozova:2018glm}. The f mode thus obtained is then laid on the spectrogram and is found to contribute significantly to the strength of the GW signal after $\sim 400$ \textrm{ms}. As noted in \citep{Morozova:2018glm}, the higher-order g- or p-modes are not found to be excited in these simulations.  

\section{Spectrogram Analysis}\label{sec:spectrogram_analysis}
In this section, we describe the construction of the spectrogram of the gravitational-wave strain signal. We use the spectrogram to measure the properties of the fundamental quadrupolar $f-$mode oscillations of the protoneutron star. In particular, we are interested in measuring the frequency of the oscillations and the energy emitted in the gravitational-wave radiation. The analysis described here is for the fiducial case when the detector noise is not present. In the later sections we will discuss the effect of detector noise in the extraction of the features from the spectrogram and compare it with the output from the analysis described here. 

Following \cite{Misner:1973prb,Oohara:1997qd, Muller:2011yi, Andresen:17}, the gravitational-wave strain $\mathbf{h^{TT}_{ij}}$ for a for a source at a distance $D$ can be written as
\begin{equation}
    \mathbf{h^{TT}_{ij}} = \frac{2G}{c^4D}\frac{dq_{ij}}{dt},\label{eq:h-quadrupole}
\end{equation}
where $q_{ij}$ is the time derivative of the mass quadrupole tensor $\mathcal{Q}_{ij} = \int d^3 x \rho (x_i x_j - \frac{1}{3}r^2 \delta_{ij})$. The strain amplitudes of the two polarizations, $h_+$ and $h_{\times}$, can be obtained in the slow-motion limit from the linear combinations of the second time derivatives of the components of the transverse traceless mass quadrupole tensor $\mathcal{Q}_{ij}$. The polarization strains as observed along the line of sight $(\theta, \phi)$ are given by
\begin{equation}
    h_+ = \frac{G}{c^4D}\left(\frac{dq_{\theta\theta}}{dt} - \frac{dq_{\phi\phi}}{dt}\right),\label{eq:h-plus}
\end{equation}
\begin{equation}
    h_{\times} = \frac{2G}{c^4D}\left(\frac{dq_{\theta\phi}}{dt}\right).\label{eq:h-cross}
\end{equation}
Here, the time derivatives of the mass quadrupole in spherical coordinates, in terms of those in Cartesian coordinates, are given by
\begin{multline}
    q_{\theta\theta} = (q_{xx}\cos^2\phi + q_{yy}\sin^2\phi + 2q_{xy}\sin\phi\cos\phi)\cos^2\theta \\ + q_{zz}\sin^2\theta - 2(q_{xz}\cos\phi + q_{yz}\sin\phi)\sin\theta\cos\theta,
\end{multline}
\begin{multline}
    q_{\phi\phi} = q_{xx}\sin^2\phi + q_{yy}\cos^2\phi - q_{xy}\sin\phi\cos\phi,
\end{multline}
\begin{multline}
    q_{\theta\theta} = (q_{xx} - q_{yy}) \cos\theta\sin\phi\cos\phi + \\ q_{xy}\cos\theta(\cos^2\phi - \sin^2\phi)\\
    + q_{xz}\sin\theta\sin\phi - q_{yz}\sin\theta\cos\phi.
\end{multline}

The total energy emitted in gravitational waves is given by \cite{Muller:2011yi}
\begin{equation}
    E_{GW} = \frac{c^3}{5G} \int_0 ^t \sum_{ij}\left[\frac{d^3 \mathcal{Q}_{ij}}{dt^3} \right]^2dt,\label{eq:energy_td_quadrupole}
\end{equation}
which, in terms of gravitational-wave strain is given by
\begin{equation}\label{eq:energy_td_strain}
\begin{split}
    E_{GW} &= \frac{c^3D^2}{16\pi G} \int_0 ^t dt \int_{4\pi} d\Omega \left[ \left(\frac{dh_+}{dt}\right)^2 + \left(\frac{dh_{\times}}{dt}\right)^2\right]\\
    &\approx \frac{c^3D^2}{4G} \int_0 ^t dt \left[ \left(\frac{dh_+}{dt}\right)^2 + \left(\frac{dh_{\times}}{dt}\right)^2\right], 
\end{split}  
\end{equation}
where the second approximation holds true if the strains are assumed to be nearly independent of line of observation and the integral over the solid angle gives a factor of $4\pi$. The variation in the amplitudes of the strain for different points of observation are $10-15\%$ \cite{Vartanyan:2019ssu}. For two-dimensional simulations without progenitor core rotation, the approximation does not hold, and we use equations 3-6 in Morozova \textit{et al} \cite{Morozova:2018glm} to compute the energy from gravitational-wave strain $h_+$. 

Ground-based interferometers will detect a linear combination of the strain polarizations,
\begin{equation}
    s(t) = s_{\mathrm{eff}}(t) + n(t),
\end{equation}
where n is the noise in the detector, and effective strain from the astrophysical source is given by,
\begin{equation}
    s_{\mathrm{eff}} = F_+ h_+ + F_{\times}h_{\times}.
\end{equation}
Here, $F_+$ and $F_{\times}$ are the antenna pattern functions of an interferometric detector, and they depend on the sky location (Right ascension, Declination) of the source at a given time, and its polarization angle with respect to the detector arms. The energy estimation from an observation is then given by
\begin{equation}\label{eq:energy_td_eff_strain}
    E_{GW; \mathrm{eff}} \approx \frac{c^3D^2}{4G} \int_0 ^t dt  \left(\frac{ds_{\mathrm{eff}}}{dt}\right)^2,
\end{equation}
which would be $\approx 0.5$ times the energy calculated using both the polarizations individually (i.e. from Equation \ref{eq:energy_td_strain}).

To compute the energy spectra of the gravitational wave signal we use the spectrogram of the signal

\begin{equation}\label{eq:energy-spectrum}
\begin{split}
    \frac{dE^{\star} _{GW}}{df} &= \frac{c^3 D^2}{4G} (2 \pi f)^2 \left[(\Tilde{h}_+)^2 + (\Tilde{h}_{\times})^2\right] \\
    &\approx \frac{c^3 D^2}{2G} (2 \pi f)^2 \left[\Tilde{s}_{\mathrm{eff}}\right]^2,
\end{split}
\end{equation}

where $\Tilde{h}$ is the short-time Fourier transform, defined as
\begin{equation}
    \Tilde{h}(f) = \int _{-\infty}^{\infty} h(t) H(t-\tau_l)e^{-2\pi ift}dt, \label{eq:stft}
\end{equation}

and $H(t-\tau_l)$ is the Hann window with offset time $\tau_l$, indexed by $l$. A window function is applied to each segment to ensure that we don't get Gibb's junk when we take the Fourier transform of the segment. Equation \ref{eq:energy-spectrum} gives the energy per unit Hertz for a time-frequency block centered at time $\tau_l$ and frequency $f$. Thus, a spectrogram is the transformation of the short-time Fourier transform to represent the power content in a time-frequency block.

\begin{figure}[t]
\includegraphics[width=\columnwidth]{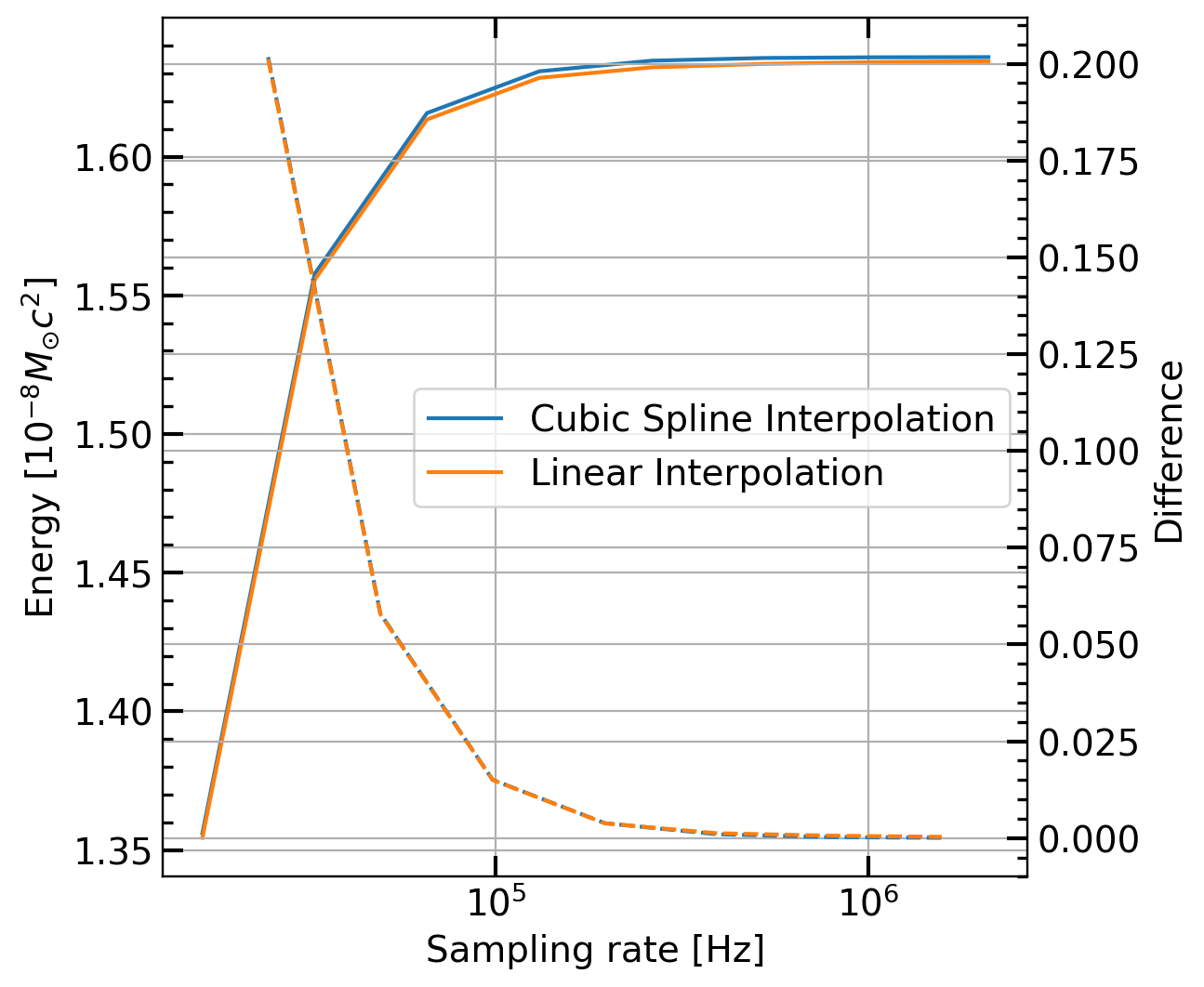}
\caption{The figure shows the energy (in $10^{-8} M_{\odot}c^2$) obtained for the simulation \texttt{s19-3D} as a function of the sampling rate used to resample the data from the simulations. The solid blue curve represent the case when Cubic interpolation is used whereas the orange curve shows the case when Linear interpolation is used. The corresponding dashed curves show the difference between energy values obtained for a particular sampling frequency and the one lower. 
We can see that the values converge as the sampling frequency is increased. 
\label{fig:interpolation-errors}}
\end{figure}

In practice, gravitational-wave strain $h$ is a discrete function of time, obtained either from simulations, or via observations made by a gravitational-wave detector. In order to take the Fourier transform of the time-domain strain data, $h_j \equiv h_+(t_j)$, it must be evenly sampled at time intervals of $\Delta t = t_{j+1} - t_j$ seconds, $\forall j$. The sampling rate, or sampling frequency is given by $f_s = 1/\Delta t$. 
The data from the simulations is unevenly sampled since the size of each time step in the simulations is governed by the micro- and macro-physics at the time. 

We re-sample the data at sampling rates ranging from $16,384$ Hz to $ 2,097,152$ Hz in powers of two and interpolate using one of the two interpolation schemes: linear interpolation and cubic spline interpolation. We then compute the energy using discretized versions of equations \ref{eq:h-plus}, \ref{eq:h-cross} and \ref{eq:energy_td_strain}, where now $h(t) \equiv h(t_j)$. We compute the third order time derivative of the the quadruple moment from the second order derivative using the central difference method. Fig.~\ref{fig:interpolation-errors} shows the energies on the left ordinate for the model \texttt{s19-3D} computed via the two interpolation methods at various sampling rates. We see that the energy values converge with increasing sampling rate. The dashed curves show the difference between the energy values obtained between two consecutive sampling rates (shown on the right ordinate). This plot gives us a range of energy estimations for data sampled at different frequencies. We choose to use the Cubic spline interpolation and a sampling rate of $16,384$ Hz (or equivalently, sampling interval of $\Delta t = 6.1035\times 10^{-5}$ seconds) since its is computationally less expensive and is a more realistic choice with regards to the sampling rate used by current and proposed gravitational-wave detectors.

In the next subsection we describe the construction of the short-time Fourier transform of the discretely-sampled signal, and measurement of the frequencies associated with the $f-$mode from the time-frequency representation. For this purpose, we use $50\%$ overlap of Hann-windowed time-segments, since this configuration does not affect the amplitude of the signal. In the subsection that follows, we discuss the construction of spectrogram that can be used to measure the energy associated with the $f-$mode oscillations. For this, we use $66.65\%$ overlap between two consecutive Hann-windowed segments since this configuration provides equal weights across all the point in the signal for power calculation.

\subsection{$f-$mode frequency measurement}

In order to compute the short-time Fourier transform of the data, we need to divide the data into segments of equal length, say of $T_{W} = N_{W} \Delta t $ seconds, and multiply each of these segments with a window function before we take its Fourier transform. There are a variety of windows available for this purpose \cite{heinzelSpectrumSpectralDensity2002}. In this study, we use the Hann window. We need to ensure that each data point of the waveform is equally weighted when we consider the sums of the windowed waveform segments. This presents a problem at the ends of the waveform, since Hann window starts from (or tapers to) zero. The solution is to first taper both ends of the waveform to zero, and then zero-pad the entire waveform on both ends by multiples of $N_{W}$ points. Zero-padding the waveform does not change the total power content in the signal since we are only adding zeros to the ends of the data. We use the window size of $T_{W} = 40$ ms. For tapering, we use the 1024 data points at both ends of the waveform and apply a half cosine window. We zero-pad both ends of the tapered waveform by $N_{W} = 655$ points. For constructing the short-time Fourier transform of the signal, we use $50\%$ overlap between two consecutive time segments that get multiplied by the Hann window.

\begin{figure}[t!]
\includegraphics[width=\columnwidth]{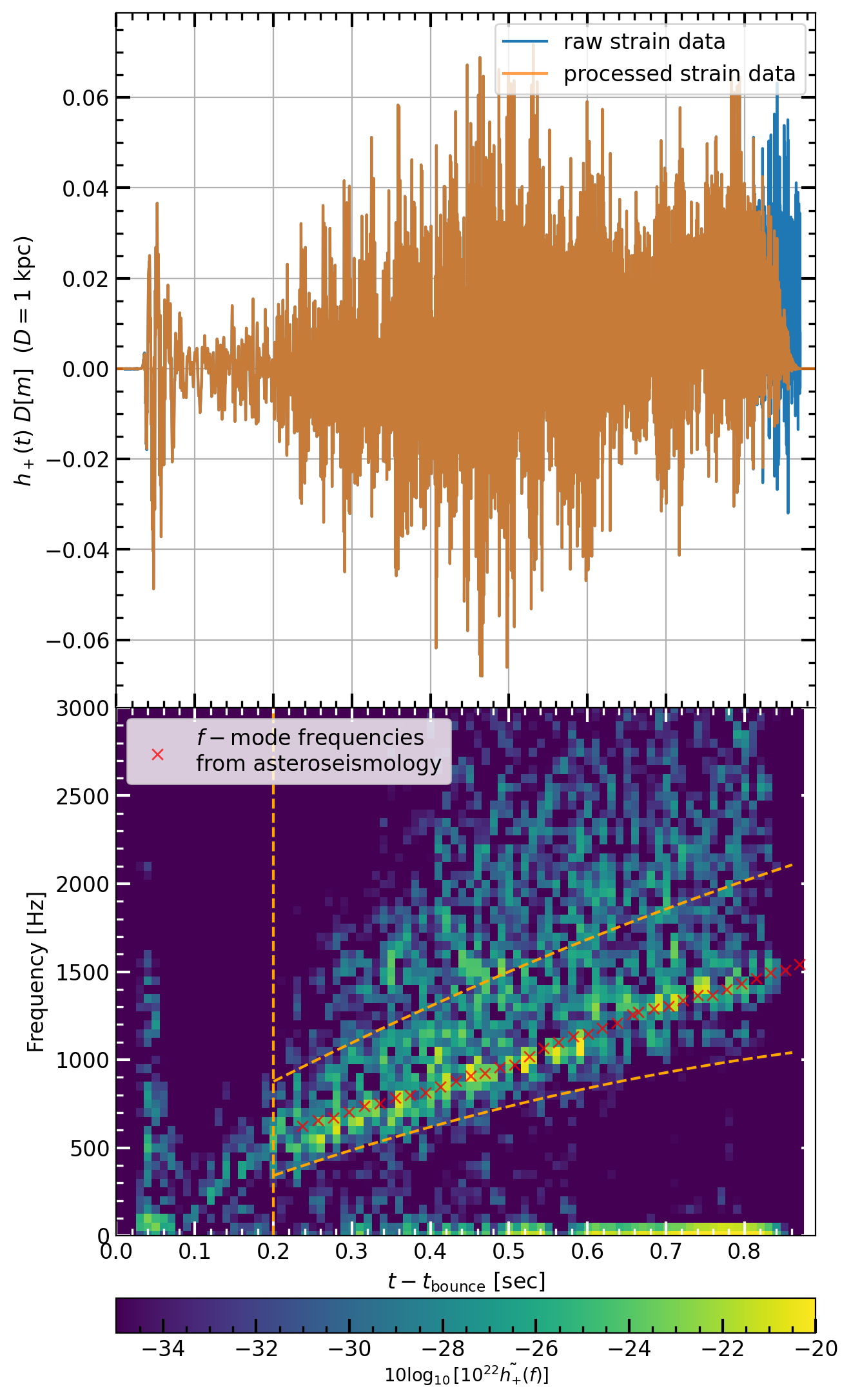}
\caption{The figure shows the strain (top) and its spectrogram (bottom) of the 3D simulation \texttt{s19-3D}. The strain data has been uniformly sampled at 16,384 Hz. The ends have been tapered to zero by applying half cosine windows to the first and last 1024 points of the data. Then, the data are zero-padded by 0.04 seconds on either end. The spectrogram of this signal is shown on the bottom panel. The $f-$mode frequencies, obtained from linear perturbation analysis, start at ~200ms and go from ~500 Hz to ~1000 Hz at 0.6 seconds after the core bounce, and are shown as red crosses in the bottom panel. The vertical orange line shows the time $t_0 = 200$ ms after the core bounce. The two orange parabolic curves define the frequency range within which the algorithm looks for the peak in the spectra.
\label{fig:s19_3D_strain_spectrogram}}
\end{figure}

The top panel of Fig.~\ref{fig:s19_3D_strain_spectrogram} shows the gravitational-wave strain data of the plus polarization as a function of time after core bounce for the simulation \texttt{s19-3D} in blue. This three-dimensional simulation uses a progenitor with ZAMS mass of $19 M_{\odot}$. The equation of state used in the simulation is SFHo. The top panel shows evenly-sampled data in blue, and the data with both ends tapered for construction of the short-time Fourier transform is shown in orange. The bottom panel shows the short-time Fourier transform of the strain. The horizontal axis shows the time after bounce, the vertical axis shows the frequency. The color bar shows the modulus of the Fourier amplitude. We see the prompt convection signal after $\sim 50$ ms after the core bounce. The prompt convection phase is followed by the $\sim 50$ ms long quiescent phase. After this, the dominant part of the signal starts with frequency growing from $\sim500$ Hz to $\sim 1000$ Hz  $0.6$ sec after core bounce. This signal is caused by matter accreting on the proto-neutron star and exciting its modes, including the $f-$mode. The $f-$mode frequencies obtained by the linear perturbation analysis are shown as red crosses in the bottom panel. 

We measure the frequency evolution of the dominant track in the short time Fourier transform. To measure the frequencies, we use the following procedure. We start by analyzing the spectrogram data after $t_0$ seconds. The vertical orange line in Fig. \ref{fig:s19_3D_strain_spectrogram} shows the time $t_0 = 200$ ms after the core bounce. From the linear perturbation analysis we know that the $f-$mode starts around this time. We define a plausible range of the $f-$mode frequencies shown by the two orange quadratic curves in Fig. \ref{fig:s19_3D_noiseless_spectrogram_analysis}.  The quadratic parameters for the lower frequency bound are  
$a = -700, b = 1800, c = 10$ and for the upper frequency bound are $a = -600, b = 2500, c = 400$ used in the formula $f(t) = a t^2 + b t + c$. For each time segment after $t_0$, we find the highest value of the energy spectrum within the frequency range constrained by the two orange curves. We model the frequency evolution of the $f-$mode as the quadratic function, and use a robust least-squares fit of the maxima in the STFT for each time slice using the \texttt{soft-l1} loss function to get the parameters $\{a, b, c\}$ of the quadratic function.

\subsection{Energy measurement}

We are interested in computing the power in each of the time-frequency blocks in the spectrogram associated with the $f-$mode frequencies. To ensure that we can do this correctly, we first compare the power in the entire signal evaluated using the Equation \ref{eq:energy_td_strain} (using the time-domain representation of the signal) and via the spectrogram (adding up power in all the time-frequency blocks). However, when we construct the spectrogram, multiplying a data segment with a window function alters the amplitude, and hence, the power, of the signal. To mitigate this problem, we make two consecutive segments of the data overlap by a fixed amount of $T_{\mathcal{O}}$ seconds. We also want to ensure that the relative weighing is the same for all the data points across different segments. The relative weighing of the data for power calculation is obtained by summing the square of the window values at each data point. That is, for the point $j$, the weighing will be given by $\sum_{l}H^2(j\Delta t-\tau_l)$. We want this quantity to be constant across the the entire signal. For the first half of the first window, and the second half of the last window, the relative weighing does not matter since we are zero-padding the ends of the waveform. 

Trethewey (2002) \cite{2000MSSP...14..267T} show that one cannot simultaneously obtain equal weighing of all data points, and compute the correct power. For Hann windows, one obtains the correct value of average power of the entire signal if consecutive segments overlap by $62.5\%$. However, in this case, there is variation in the relative weighing of the data points, resulting in amplification of power in certain data segments, whereas reduction in others. This would mean large errors in power estimates within individual time segments, specially if the data are stochastic in nature, like the gravitational-wave strain from a supernova. If two consecutive segments overlap by $66.65\%$, all the points are equally weighed, but the power calculation is amplified by a factor of $1.125$, across all segments. However, we can compute the power with $66.65\%$ overlap, and scale it down by the relevant factor to obtain the correct value of power.  

Assuming $66.65\%$ overlap between segments, the time-difference between the start of two consecutive segments is $T_{H} = T_{W} - T_{\mathcal{O}}$ seconds, such that $100 \times T_{\mathcal{O}}/T_{W} = 66.65$. We multiply each segment by a Hann window $H(t_j -\tau_l)$, where $\tau_l$ is the time offset of the center of the segment from the start of the signal. The length of the Hann window is equal to the length of the segment. We take the discrete Fourier transform of each segment of the windowed data $h_jH(t_j - \tau_l)$ using \texttt{scipy.fftpack.fft}, given by
\begin{equation}\label{eq:numrec_DFT}
\tilde{y}_{k,\tau_l} = \sum_{j=0}^{N_{W}-1} h_jH(t_j - \tau_l) e^{2\pi ijk/N_{W}}.
\end{equation}


The discrete form of Equations \ref{eq:energy-spectrum}, \ref{eq:stft} is given by 
\begin{equation}\label{eq:dEbydf_descrete}
 \frac{\Delta E_{GW}}{\Delta f}(f_k, \tau_j) \sim \frac{D^2c^3}{2G}(2\pi f_k)^2 \Delta t^2 |\tilde{y}_{k,\tau_j}|^2,
\end{equation}
where we have a factor of 2 instead of 4 in the denominator to account for the power in the negative frequencies. To normalize the effect of the window function, we use the window normalizatoin factor from Heinzel et al (2002)  \cite{heinzelSpectrumSpectralDensity2002},\begin{equation}\label{eq:s2_definition}
\frac{S_2}{N_{W}} = \frac{1}{N_{W}}\sum_{j=0}^{N_{w}-1} H^2(t_j).
\end{equation}
Using this normalization, Equation \ref{eq:dEbydf_descrete} becomes
\begin{equation}\label{eq:dEbydf_descrete_normlaized}
 \frac{\Delta E_{GW}}{\Delta f}(f_k, \tau_j) = \frac{D^2c^3}{2G}(2\pi f_k)^2 \Delta t^2 |\tilde{y}_{k,\tau_j}|^2\left[\frac{N_{W}}{S_2}\right].
\end{equation}
The total Energy can then be written as
\begin{equation}\label{eq:total_energy}
\begin{split}
    E_{GW} & = \sum_{k}\sum_{l} \frac{\Delta E_{GW}}{\Delta f}(f_k, \tau_j) \Delta f \left[\frac{T_{H}}{T_{W}}\right] \\
    & = \sum_{k}\sum_{l} \frac{D^2c^3}{2G}(2\pi f_k)^2 \Delta t^2 |\tilde{y}_{k,\tau_j}|^2 \Delta f \left[\frac{N_{W} T_{H}}{S_2 T_{W}}\right],
\end{split}
\end{equation}
where the quantity $\left[\frac{T_{H}}{T_{W}}\right]$ is introduced to account for the fact that two consecutive segments overlap by $T_{\mathcal{O}}$ seconds. The factor $\left[\frac{T_{H}}{T_{W}}\right]$  corrects for this double-counting and represents the correct fraction of the energy in the time segment. in the final equation, the factor $\left[\frac{N_{W} T_{H}}{S_2 T_{W}}\right]$ for $66.65\%$ overlap is equal to $1.125$, the same factor from Trethewey (2000) \cite{2000MSSP...14..267T}.

\begin{figure}[t!]
\includegraphics[width=\columnwidth]{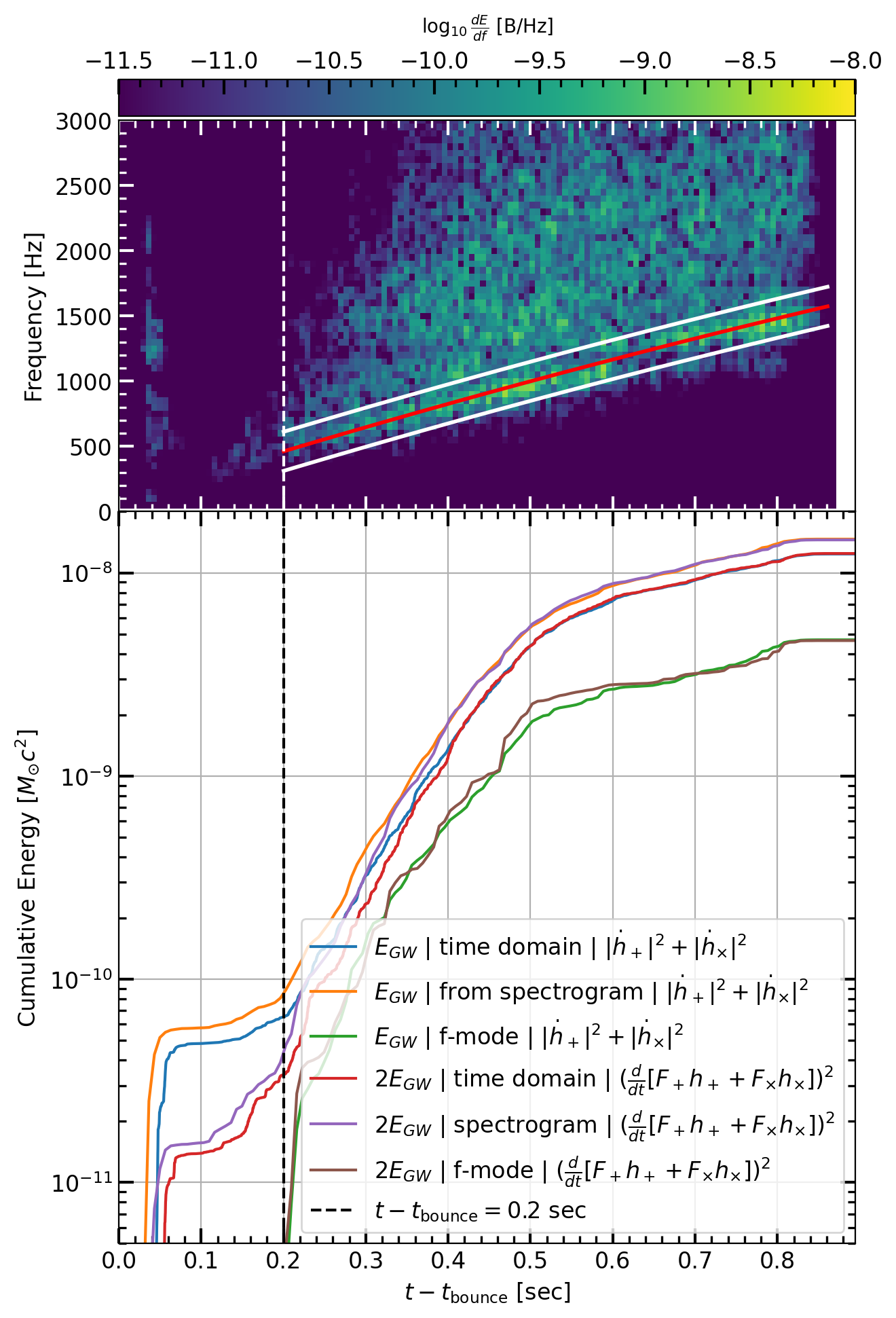}
\caption{The top panel in the figure shows the spectrogram of the 3D simulation \texttt{s19-3D}. The red curve shows the quadratic fit to the fundamental $f-$mode frequencies measured by picking the frequencies corresponding to the peak in spectra and the two white curves represent the width of the track. The energy of the mode is measured by summing the energy of the time-frequency blocks within the white curves. The blue, orange, and green curves in the bottom panel show the cumulative energy measured from the time domain representation of the signal, its spectrogram, and for the $f-$mode from the spectrogram, respectively, and summed for the individual polarizations $h_+$ and $h_{\times}$. The red, purple, and brown curves show two times the energies for the same for the effective strain as observed in a interferometer. 
\label{fig:s19_3D_noiseless_spectrogram_analysis}}
\end{figure}

Now, we can verify if the energy values obtained from the spectrogram (Equation \ref{eq:total_energy}) agree with those obtained using the Equation \ref{eq:energy_td_strain}. We find that the energy values agree within the error range due to interpolation. 

Once we have verified that the energies obtained from the time-frequency data agree with those obtained by the time domain data, we can compute the energy associated with a given time-frequency track. In particular, we can compute the energy associated with the $f-$mode oscillation of the protoneutron star. 

The top panel in Fig. \ref{fig:s19_3D_noiseless_spectrogram_analysis} shows the frequency evolution of the $f-$mode obtained from the spectrogram. For each time-segment, we assume the width of the $f-$mode track to be $6\Delta f$ ($3\Delta f$ above the spectral peak associated with the $f-$mode frequencies, and $3\Delta f$ below it). This width is represented as the two white curves encompassing the peak frequency curve shown in red. We can add up the energy values for all the time-frequency blocks within the width obtained. Doing this for all the time segments after $t=t_0$ will give us the time evolution of the energy associated with the $f-$mode. 

The bottom panel of Fig. \ref{fig:s19_3D_noiseless_spectrogram_analysis} shows the cumulative energy as a function of time for the simulation \texttt{s19-3D}. The blue curve shows the cumulative energies obtained from the time-domain data (Equation \ref{eq:energy_td_strain}) and the orange curve shows the cumulative energies obtained by adding up energy values for all time-frequency blocks in the spectrogram. We can see that both match very well. The green curve shows the cumulative energy of the $f-$mode as a function of time measured by adding the energy values in time-frequency blocks only corresponding to the $f-$mode (i.e. within the two white curves in top panel of Fig. \ref{fig:s19_3D_noiseless_spectrogram_analysis}). The energy obtained in the $f-$mode is $\approx 20-40\%$ of the energy from the entire signal.  We also compute the energy values obtained from the effective strain observed by a detector and its spectrogram (i.e. from Equation \ref{eq:energy_td_eff_strain}). We show these energies (multiplied by a factor of 2) in purple and red color in the bottom panel.  

We test this method for toy signals of the form $h(t) = A(t)\sin(2\pi f t)$, as well as the solutions for the differential equations for driven simple-harmonic oscillator (for details on this model see Ref. \cite{Astone:2018uge}). We find that we get $\approx 20\%$ error in the measurement of the power from the spectrogram as compared to the measurement directly from the time domain signal. The error is higher when stochasticity of the amplitude increases.

\section{Results}\label{sec:results}

In this section, we describe the results of using the above method for measuring the properties of the $f-$mode for gravitational-wave strains from various simulations. 

\begin{figure*}[t]
\includegraphics[width=18cm]{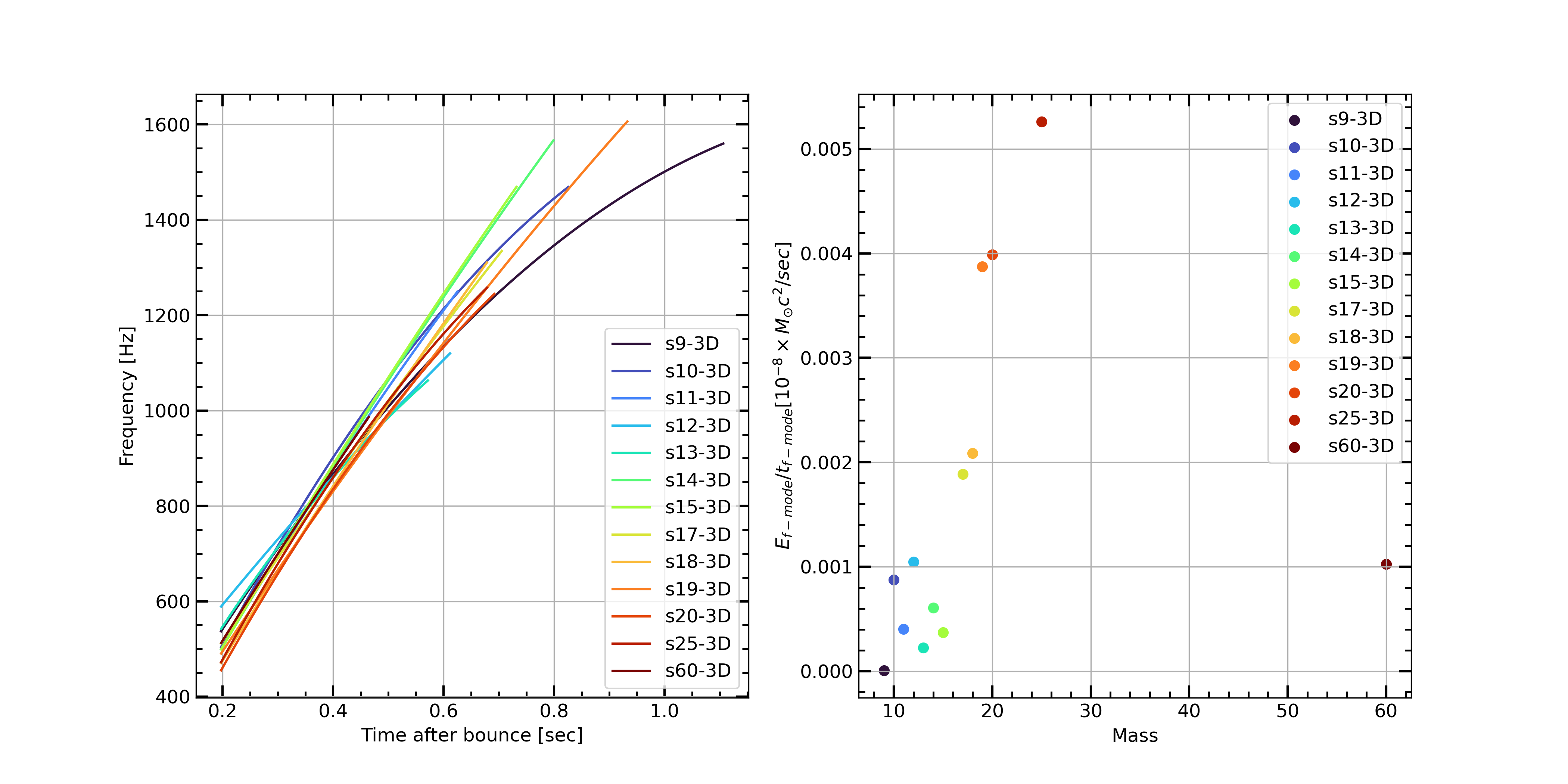}
\caption{The figure shows the frequency evolution (left) and power (right) measured for the $f-$mode oscillations from the gravitational-wave strains of the three-dimensional simulations. The frequency increases with time, owing to the shrinking of the proto-neutron star. We find that the power in the gravitational-waves associated with the $f-$mode oscillations generally increases as the progenitor mass increases.   
\label{fig:3D_waveforms_fmode_freqs_energy}}
\end{figure*}

The left panel of Fig.~\ref{fig:3D_waveforms_fmode_freqs_energy} shows the frequency evolution of the $f-$mode measured from short-time Fourier transform for the three-dimensional simulations. Since the simulations are for a short duration, we start the measurement of the $f-$mode frequencies (and consequently, the energy) from $200$ ms after the core bounce. However, this procedure makes the frequency measurement noisy for the time interval $200 - 400$ ms after the core bounce since $f-$mode oscillations are not the most energetic contributors to the gravitational-wave signal. The peak of the Fourier transform may not lie on the frequencies associated with the $f-$mode. One can see that the frequencies at $200$ ms lie in the range between $500$ Hz - $600$ Hz. $600$ ms after the core bounce, the frequencies can increase up to $1100$ Hz - $1250$ Hz. Again, there is no monotonous dependence of the frequencies with the progenitor mass. 

The right panel of Fig.~\ref{fig:3D_waveforms_fmode_freqs_energy} shows the energy in the $f-$mode track obtained from the spectrogram of the gravitational-wave signal, divided by the time when the $f-$mode oscillations are active (i.e. 200 ms to end of simulations). Typically, the higher mass progenitors produce a stronger gravitational-wave signal, and hence the power in the $f-$mode oscillation track is higher for higher progenitor mass. If we look at Fig.~2 of Ref \cite{Burrows:2019zce}, which describes the time evolution of the shock radius of the same progenitors that we use in this study, we find that delayed explosion time also correlates with increased energy emission in the gravitational wave signal. For example, the power measured in $f-$mode for simulations with progenitor masses $10, 19,$ and $25 M_{\odot}$ is higher in comparison to other simulations, and the shock expansion associated with the explosion is also delayed. For the progenitors with no explosion during the simulation time, we measure low power in the $f-$mode from the gravitational-wave signal.

\begin{figure*}[ht]
\includegraphics[width=18cm]{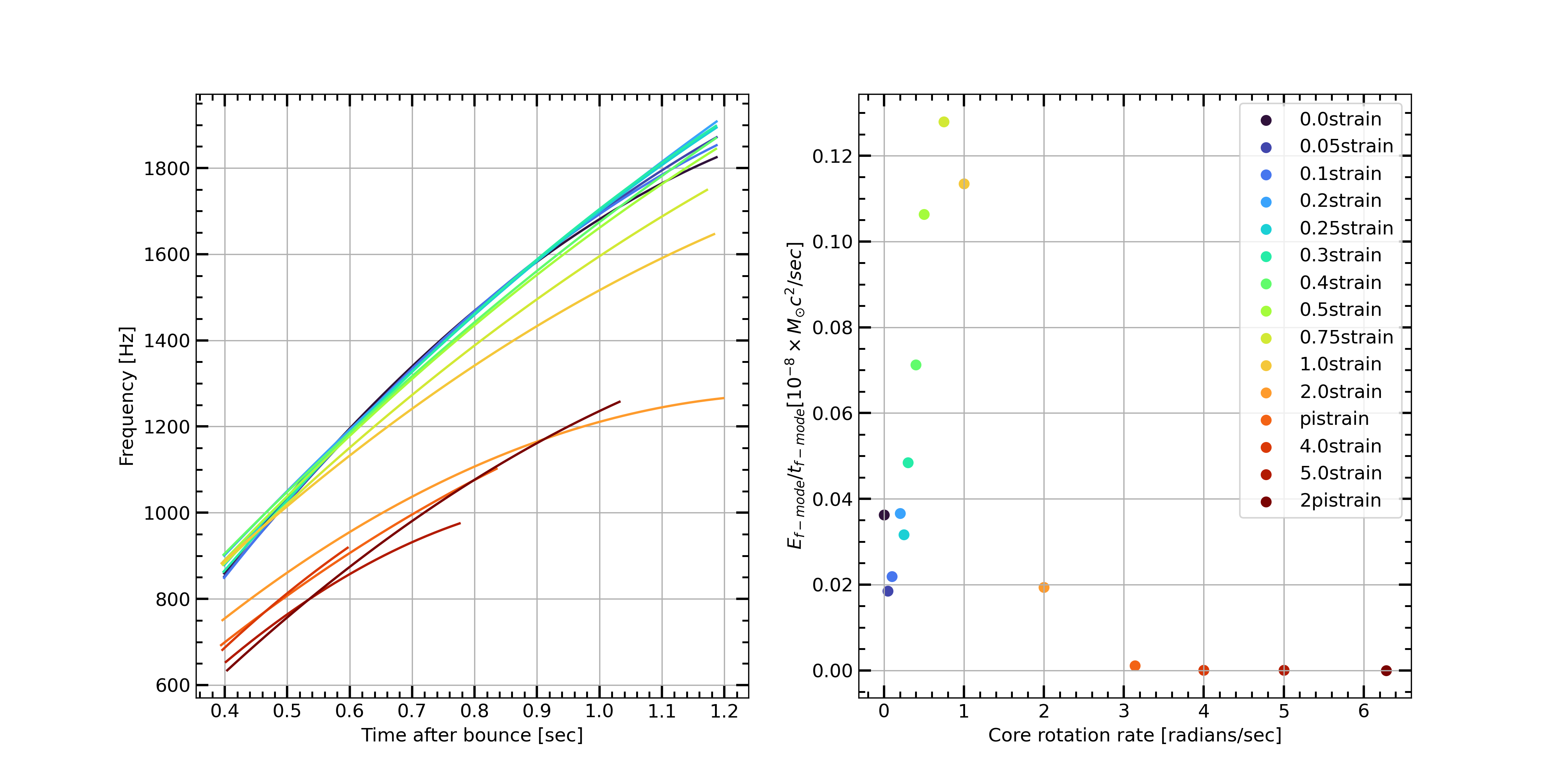}
\caption{The figure shows the frequency evolution (left) and power (right) measured for the $f-$mode oscillations from the gravitational-wave strains of the two-dimensional simulations with core rotation. The frequency increases with time, owing to the shrinking of the proto-neutron star. We find that the power in the gravitational-waves associated with the $f-$mode oscillations increases monotonically as the progenitor core rotation rate increases, up till $\Omega = 0.75$ rad/sec, and then decreases as centrifugal forces dominate. 
\label{fig:2d_rotating_waveforms_fmode_freqs_energy}}
\end{figure*}

Fig.~\ref{fig:2d_rotating_waveforms_fmode_freqs_energy} shows the frequency evolution on the left panel and the power in $f-$mode oscillations from the gravitational-wave signal as a function of the core rotation rate on the right panel for simulations with non-zero core rotation. We find that both the frequency evolution and the power of the $f-$mode oscillations depend on the core rotation rate. Mild core rotation (0 - 0.75 rad/sec) increases the quadrupole moment and hence the power in gravitational-wave radiation. Increasing the core rotation rate increases the centrifugal support on the accreting matter that excites the proto-neutron star oscillations, resulting in reduced power in the oscillations. The frequencies are also affected and we can see two distinct groups of the frequency tracks. For the core rotation rates of $0 - 0.5$ rad/sec, where the centrifugal forces are not affecting the $f-$mode oscillations, we see we see similar time dependence for the frequencies as for the simulations with no core rotation --- they rise from $\approx 850 - 900$ Hz at $400$ ms after the core bounce to $\approx 1700$ Hz at 1 second after the core bounce. As the core rotation rate increases, the frequencies decrease, as is seen in the simulations with $0.75$ rad/sec and $1.0$ rad/sec core rotation rates. Further, for the higher rotation rates (greater than $1.0$ rad/sec), the centrifugal forces are large and the frequencies decrease - starting from $\approx 600 - 800$ Hz at $400$ ms after the core bounce to $\approx 1200$ Hz at 1 second after the core bounce.

\begin{figure*}[ht]
\includegraphics[width=18cm]{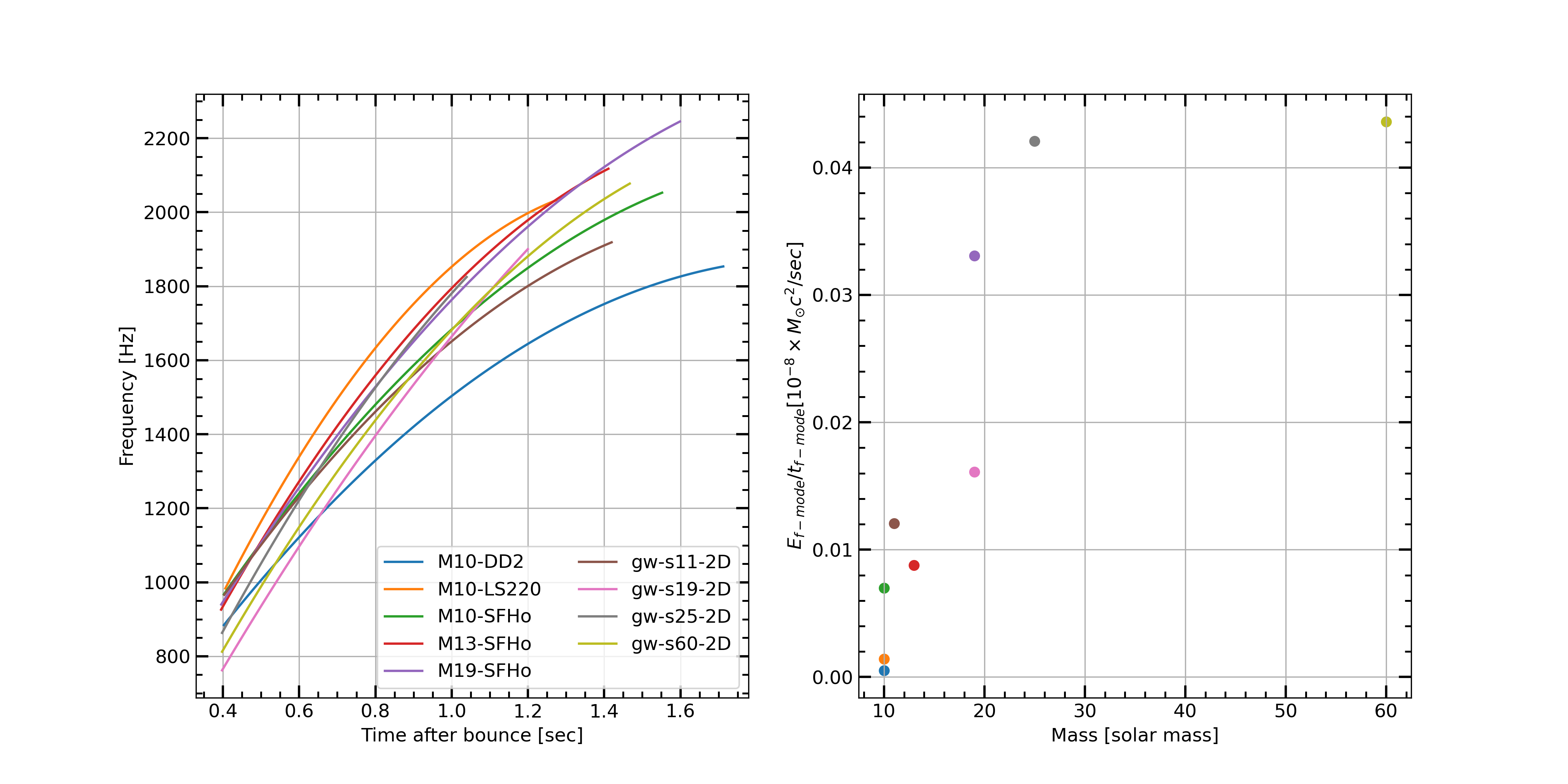}
\caption{The figure shows the frequency evolution (left) and power (right) measured for the $f-$mode oscillations from the gravitational-wave strains of the two-dimensional simulations without core rotation. The frequency increases with time, owing to the shrinking of the proto-neutron star. We find that the power in the gravitational-waves associated with the $f-$mode oscillations generally increases as the progenitor mass increases.  
\label{fig:2d_nonrotating_waveforms_fmode_freqs_energy}}
\end{figure*}

The left panel of Fig.~\ref{fig:2d_nonrotating_waveforms_fmode_freqs_energy} shows the interpolated $f-$mode frequency evolution measured from the short-time Fourier transforms of the simulations listed in the figure. These are two-dimensional simulations with progenitors having zero core rotation rate at the core bounce. We assume that the $f-$mode starts at $\sim 400$ ms after the core bounce.  There is no monotonic trend with respect to the mass of the progenitor star for the frequency evolution of the $f-$mode oscillations. There is also no monotonic dependence on the equation of state used in the simulation. The stiffest equation of state, DD2, used for the simulation \texttt{M10-DD2} produces the smallest frequencies (for times $\sim 600$ ms after the core bounce). Whereas, the softest equation of state, SFHo, used for simulation \texttt{M10-SFHo}, produces frequencies lower than a relatively stiffer equation of state LS220. This has been already discussed in Morozova \textit{et al.} \cite{Morozova:2018glm}, where the authors obtained the frequency evolution from linear perturbation analysis of the proto-neutron star. Here we verify the frequency evolution by measuring the frequencies from the short-time Fourier transform. The right panel of Figure~\ref{fig:2d_nonrotating_waveforms_fmode_freqs_energy} shows the power in the $f-$mode track obtained from the spectrogram of the gravitational-wave signal. Again we see that the higher mass progenitors typically have higher power in the $f-$mode oscillation track. The results of energy measurement for all the waveforms are summarized in Table~\ref{tab:results}.

\hspace{-1cm}
\begin{table}
\vspace{2cm}
\centering
\begin{tabular}{|c|c|c|c|c|}
\hline
\hline
\multirow{2}{*}{Label}&\multicolumn{4}{c|}{$E_{GW}(10^{-8} M_\odot c^2)$} \\
\cline{2-5}
 &  \makecell{Time\\Domain} & \shortstack[c]{Spectrogram}  & \shortstack[c]{Fractional \\ Error} &\shortstack[c]{$f-$mode \\energy} \\
\hline

\texttt{s9-3D}   & 0.0014 & 0.0015  & 7 $\%$  & 0.0002  \\
\texttt{s10-3D}  & 0.073 & 0.084  & 15 $\%$ & 0.016  \\
\texttt{s11-3D}  & 0.03 & 0.03  & 6 $\%$  & 0.005   \\
\texttt{s12-3D}  & 0.05 & 0.05 & 1 $\%$ & 0.012  \\
\texttt{s13-3D}  & 0.012 & 0.011  & 9 $\%$  & 0.002  \\           
\texttt{s14-3D}  & 0.028 & 0.031  &  10 $\%$ & 0.003  \\
\texttt{s15-3D}  & 0.026 & 0.029  & 8 $\%$  & 0.006  \\
\texttt{s17-3D}  & 0.12 & 0.13  & 14 $\%$  & 0.03  \\
\texttt{s18-3D}  & 0.13 & 0.15 & 10 $\%$  & 0.03  \\
\texttt{s19-3D}  & 0.33 & 0.38  & 16 $\%$  & 0.081  \\
\texttt{s20-3D}  & 0.18 & 0.21  & 16 $\%$  & 0.27  \\
\texttt{s25-3D}  & 0.13 & 0.16  & 21 $\%$  & 0.07  \\
\texttt{s60-3D}  & 0.024 & 0.027  & 13 $\%$  & 0.008  \\
\hline

 
\texttt{0.0strain} & 1.2 & 1.5 & 22 $\%$  & 0.81  \\
\texttt{0.05strain} & 0.92 & 1.08 & 18 $\%$  & 0.41  \\
\texttt{0.1strain} & 0.97 & 1.15 & 19 $\%$  & 0.49  \\
\texttt{0.2strain} & 1.43 & 1.63 & 14 $\%$  & 0.82  \\ 
\texttt{0.25strain} & 1.09 & 1.34 & 22 $\%$  & 0.71 \\
\texttt{0.3strain} & 1.47 & 1.77 & 20 $\%$  & 1.09  \\ 
\texttt{0.4strain} & 2.28 & 2.92 & 28 $\%$  & 1.60  \\
\texttt{0.5strain} & 3.75 & 4.38 & 17 $\%$  & 2.38 \\
\texttt{0.75strain} & 5.63 & 6.47 & 15 $\%$  & 2.82  \\
\texttt{1.0strain} & 7.9 & 8.7 & 9.4 $\%$  & 2.54  \\
\texttt{2.0strain} & 5.7 & 6.3 & 10 $\%$  & 0.44  \\
\texttt{pi.strain} & 7.44 & 8.13 & 9.3 $\%$  & 0.01 \\
\texttt{4.0strain} & 5.07 & 4.07 & 20 $\%$  & 5e-4  \\
\texttt{5.0strain} & 2.41 & 2.56 & 6 $\%$  & 5e-5  \\
\texttt{2pistrain} & 0.96 & 3.3 & 242 $\%$  & 9e-8  \\
\hline

\texttt{M10-LS220}  & 0.21 & 0.24 & 13 $\%$ & 0.07   \\
\texttt{M10-DD2}    & 0.16 & 0.17 & 11 $\%$ & 0.04   \\
\texttt{M10-SFHo}   & 1.57 & 1.86 & 19 $\%$ & 0.43   \\
\texttt{M13-SFHo}   & 0.93 & 1.09 & 17 $\%$  & 0.47  \\
\texttt{M19-SFHo}   &  5.27 & 6.80  & 29 $\%$  & 2.12  \\
\texttt{gw-s11-2D}  & 1.54 & 1.89  & 22 $\%$  & 0.65  \\
\texttt{gw-s19-2D}  & 1.47 & 1.74  & 18 $\%$  & 0.69  \\
\texttt{gw-s25-2D}  & 4.52 & 5.46  & 20 $\%$  & 1.44  \\
\texttt{gw-s60-2D}  & 4.14 & 5.48  & 32 $\%$  & 2.49  \\

\hline
\hline
\end{tabular}
\caption{In this table we show the energy in the gravitational-wave signal computed from the time-domain representation of the signal (equation \ref{eq:energy_td_strain}), and from the spectrogram (equation \ref{eq:energy-spectrum}). We show the error in measurement of the energy from the spectrogram of the signal. In the last column, we show the energy measured from the spectrogram in the track associated with the $f-$mode oscillations.} \label{tab:results}
\end{table}

We also repeat the analysis on the spectrogram of signals embedded in simulated detector noise. We assume the source distances to range from $1$ kpc to $60$ kpc. For each distance, we inject the signal in $10000$ instances of detector noise. To generate the simulated noise instances, we use the designed power spectral density for Advanced LIGO and the proposed third generation detectors, Cosmic Explorer and Einstein Telescope. 

We assume that the time of core bounce will be measured by the neutrino detectors such as ICECUBE\cite{IceCube:2011cwc}, Super-Kamiokande \cite{Ikeda:2007sa}, and DUNE \cite{Acciarri:2015uup} to within 4 ms \cite{Tamborra:2017ull, Yokozawa:2014tca, Muller:2019upo}. We measure the $f-$mode frequencies $200$ ms after the time of core bounce. We also assume that the distance to the progenitor is known \textit{a priori}, so that when we measure the energy in the $f-$mode via the spectrogram, we can scale it (by the square of the distance) to obtain the energy in gravitational-waves associated with the $f-$mode.

For closer sources, the signal is strong and the noise does not affect the $f-$mode frequency measurement. As the source distance increases, the $f-$mode peaks are picked more randomly. This is because the gravitational-wave strain amplitude from the signal is dominated by the detector noise. Consequently, the least-square fit is also affected.

\begin{figure*}
\vspace{-2.5cm}
\includegraphics[width=0.9\textwidth]{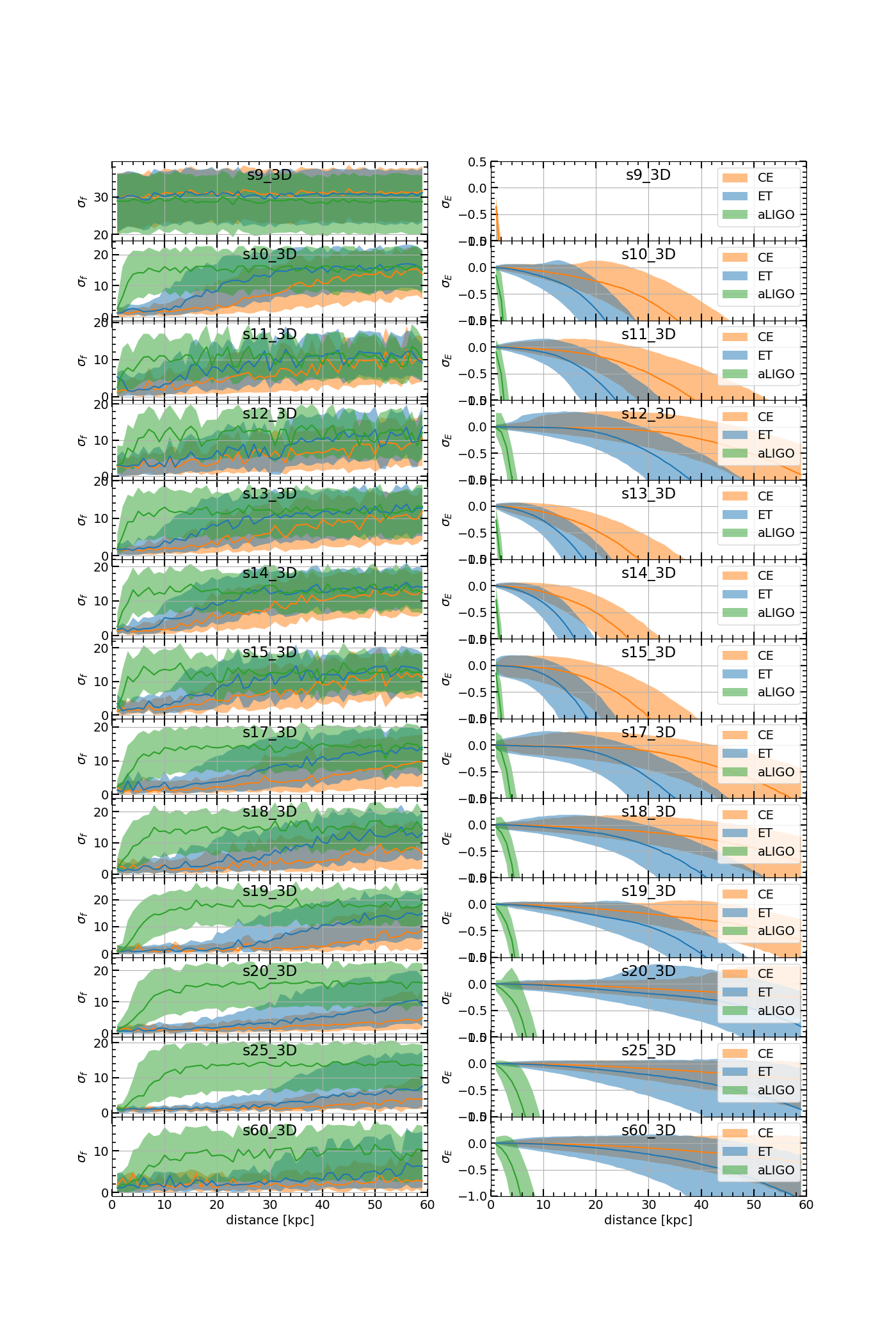}
\vspace{-2.5cm}
\caption{The left panels of the figure show the root-mean-squared error in measurement of frequency evolution of f-mode ($\sigma_f$) for waveforms from three-dimensional simulations. The right panel shows the error in measurement of energy in the $f-$mode oscillations ($\sigma_E$). The orange line shows the median obtained from measurement in 10000 noise instances of Cosmic Explorer noise, with the fill representing the 90th quantile. The blue curve represents the results for Einstein Telescope, and the green curve for Advanced LIGO.\label{fig:3D_nonrotating_core_simulation_fmode_measurement}}
\end{figure*}

\begin{figure*}
\vspace{-2.5cm}
\includegraphics[width=0.9\textwidth]{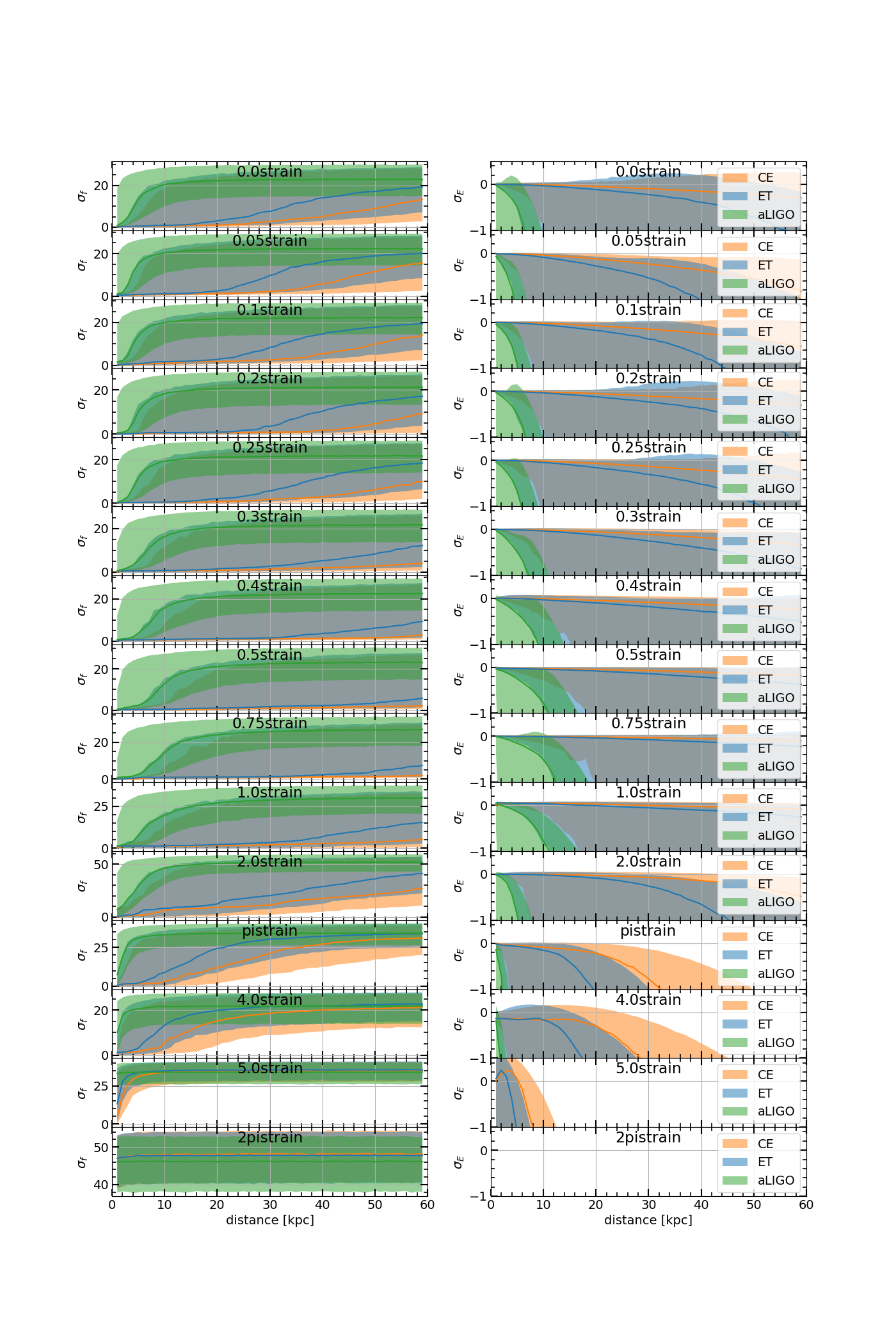}
\vspace{-2.5cm}
\caption{The left panels of the figure show the root-mean-squared error in measurement of frequency evolution of f-mode ($\sigma_f$) for waveforms from two-dimensional simulations with core rotation. The right panel shows the error in measurement of energy in the $f-$mode oscillations ($\sigma_E$). The orange line shows the median obtained from measurement in $10000$ noise instances of Cosmic Explorer noise, with the fill representing the $90$th quantile. The blue curve represents the results for Einstein Telescope, and the green curve for Advanced LIGO.\label{fig:2D_rotating_core_simulation_fmode_measurement}}
\end{figure*}

\begin{figure*}
\vspace{-2.5cm}
\includegraphics[width=0.9\textwidth]{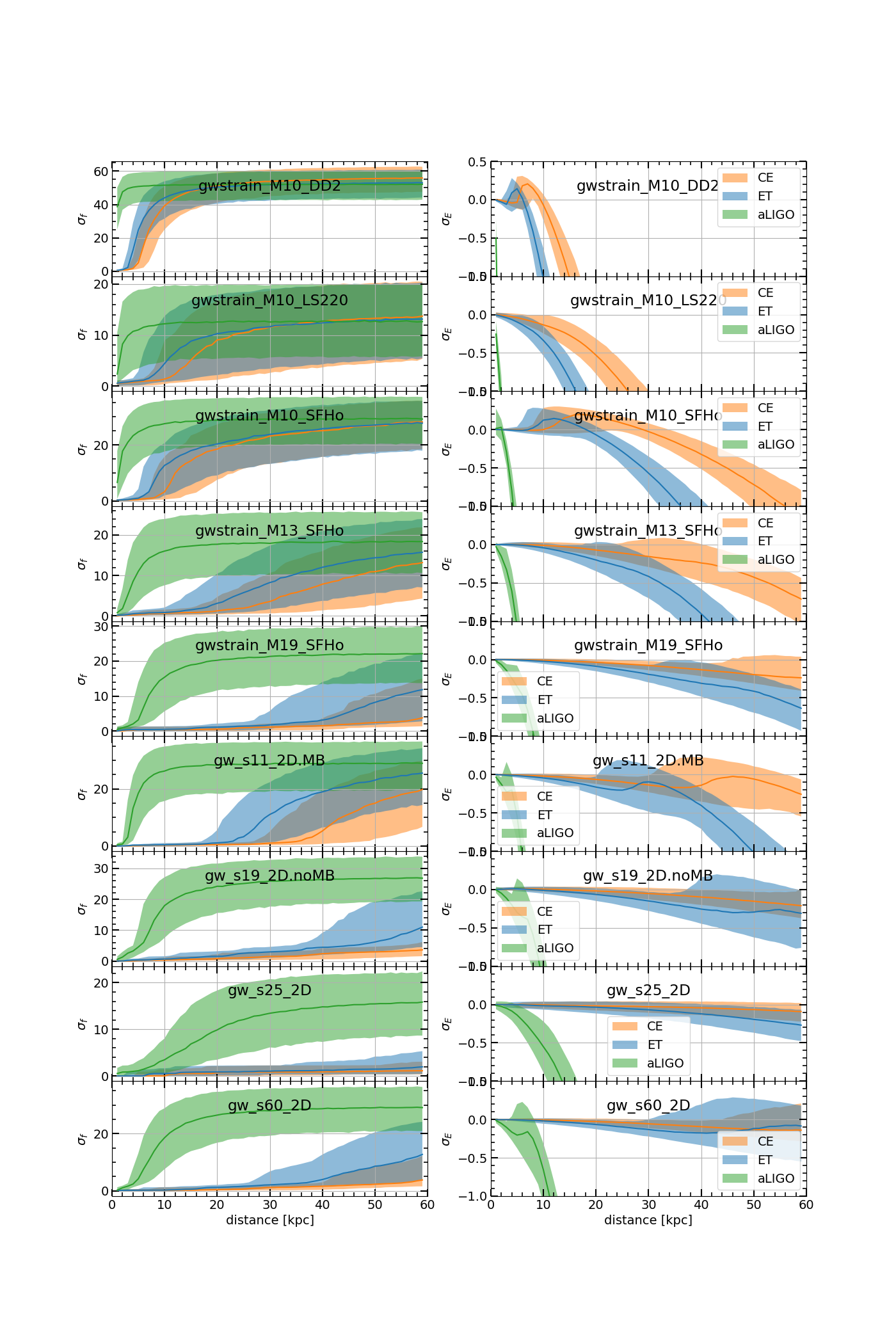}
\vspace{-2.5cm}
\caption{The left panels of the figure show the root-mean-squared error in measurement of frequency evolution of f-mode ($\sigma_f$) for waveforms from two-dimensional simulations with zero core rotation. The right panel shows the error in measurement of energy in the $f-$mode oscillations ($\sigma_E$). The orange line shows the median obtained from measurement in 10000 noise instances of Cosmic Explorer noise, with the fill representing the 90th quantile. The blue curve represents the results for Einstein Telescope, and green curve for Advanced LIGO.\label{fig:2D_nonrotating_core_simulation_fmode_measurement}}
\end{figure*}

In Figs.~\ref{fig:3D_nonrotating_core_simulation_fmode_measurement},\ref{fig:2D_rotating_core_simulation_fmode_measurement},\ref{fig:2D_nonrotating_core_simulation_fmode_measurement}, we show the results of our analysis for the $f-$mode frequency and energy measurement when the signal is embedded in detector noise. For each case, we generate a short-time Fourier transform and measure the $f-$mode frequencies. We then interpolate the measured frequencies, and compare them with those for the case when the signal was not embedded in simulated detector noise. We do this by computing the root-mean-squared error in the frequencies, given by
\begin{equation}
   \sigma_f = \sqrt{\frac{1}{N_f} \sum_1^{N_f}\Delta_f^2}.
\end{equation}
Here, $\Delta_f^2 = (f_{\mathrm{with\,noise}} - f_{\mathrm{noiseless}})^2$ and $N_f$ is the number of time columns of the $f-$mode. After measuring the $f-$mode frequencies, we use them to measure the energy in the corresponding track on the spectrogram. Since this track now has the gravitational-wave signal from a core-collapse as well as the detector noise, the measurement will yield the energy in the sum of the two. This way, we can place an upper bound on the energy in gravitational-wave radiation associated with the $f-$mode.

In Figs.~\ref{fig:3D_nonrotating_core_simulation_fmode_measurement},\ref{fig:2D_rotating_core_simulation_fmode_measurement},\ref{fig:2D_nonrotating_core_simulation_fmode_measurement}, on the left panels, we show error in frequency measurement ($\sigma_f$) on the vertical axis and distance to the source on the horizontal axis. The blue curve shows the median of the inner product obtained from 10000 injections into simulated Cosmic Explorer noise, whereas the green curve shows the same for the Einstein Telescope. The fill represents the 90th quantile measurements of $\sigma_f$. We see that as distance increases, $\sigma_f$ also increases. we can measure the $f-$mode frequencies to within $10\%$ error for sources within the Milky Way galaxy. 

The right panel shows the relative error in measurement of the energy in the $f-$mode from the spectrogram  based on the frequencies measured from the short-time Fourier transform,
\begin{equation}
   \sigma_E = \frac{E_{f-\mathrm{mode | noiseless}}- E_{f-\mathrm{mode | with\,noise}}}{E_{f-\mathrm{mode | noiseless}}}.
\end{equation}
We find that we overestimate the energy by up to $\approx 20\%$ for higher mass exploding models with source distances within the Milky Way galaxy.

\begin{figure}[ht]
\includegraphics[width=\columnwidth]{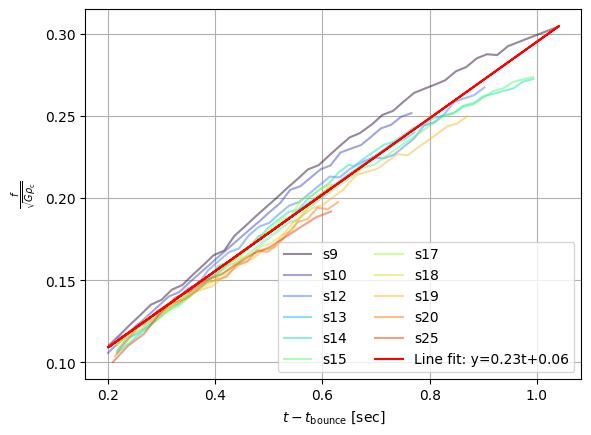}
\caption{$\frac{f_{\mathrm{f-mode}}}{\sqrt{G \rho_c}}$ as a function of time for the three-dimensional simulations, where $\rho_c$ is the central density of the protoneutron star. The frequencies are obtained from linear perturbation analysis whereas the $\rho_c$ values are obtained from the simulation data. We obtain a linear fit $\frac{f_{\mathrm{f-mode}}}{\sqrt{G \rho_c}} = 0.23t+0.06$ (shown in red) using the data for all the simulations.}\label{fig:freq-over-sqrtGrhoc-fits}
\end{figure}

\begin{figure}[ht]
\includegraphics[width=\columnwidth]{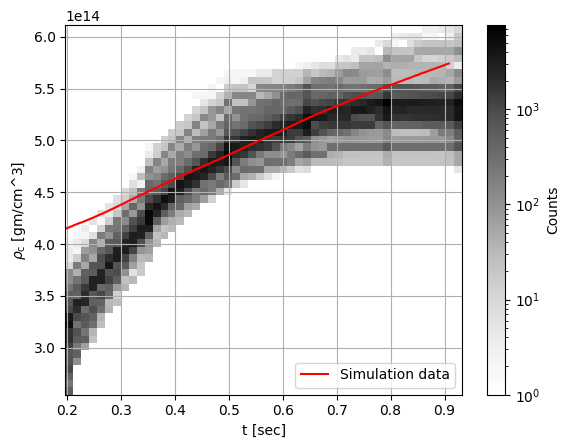}
\caption{We plot the time evolution of $\rho_c$ for the 19 $M_{\odot}$ model in red, as obtained from the three-dimensional simulation. We obtain $\rho_c$ for each injection instance when the source is assumed to be at 10 kpc and the signal is detected in Cosmic Explorer. We plot the two-dimensional histogram for the $\rho_{\mathrm{inj}}(t_b)$ where $\mathrm{inj}$ is the injection instance and $t_b$ is the time after bounce. The counts for such histogram is shown on the colorbar.}\label{fig:rho_c-recovery}
\end{figure}

In Fig. \ref{fig:freq-over-sqrtGrhoc-fits} we show the time evolution of the ratio $\frac{f_{\mathrm{f-mode}}}{\sqrt{G \rho_c}}$ for the three-dimensional simulations, where $\rho_c$ is the central density of the protoneutron star. We find that the ratio linearly increases with time, and obtain the fit $y = 0.23t + 0.06$ using linear regression, where t is the time after bounce. The fit is shown in red in Fig.~ \ref{fig:freq-over-sqrtGrhoc-fits}. Using this fit, and the frequency evolution from the spectrogram of the strain measured in a detector, we can measure $\rho_c ( t) $. In Fig. \ref{fig:rho_c-recovery} the central density $\rho_c (t) $ of the 19 $M_{\odot}$ star obtained from simulation is shown in red. We also measure $\rho_c$ for the 10000 injections of the signal associated with the model $s19-3D$ in Cosmic Explorer, assuming the source distance to be 10 kpc. Given the linear fit for $\frac{f_{\mathrm{f-mode}}}{\sqrt{G \rho_c}}$ and quadratic fits for $f(t)$, $\rho_c (t)$ is a $\sim t^2$ function of time after bounce. We obtain $\rho_c$ values for various times and for various injection instances. The two-dimensional histogram for $\rho_c$ is shown in grey-scale, with the colormap normalized to the logarithm of counts in each $\rho_c - t$ bin. We can see from the plot that we can measure the central density of the core of the star using the frequency evolution measured from the spectrogram.

\section{Conclusion}\label{sec:conclusion}

In this paper, we have developed a model-independent method to measure the frequencies and energies associated with the quadrupolar oscillations of a proto-neutron star. We use gravitational-wave signals from two- and three-dimensional core-collapse simulations. 

We construct the short-time Fourier transform of these signals to extract the $f-$mode frequencies. We then construct a spectrogram of the signal in a way that provides equal weights in power to all the data points of the signal.  

We first test the energy measurement from the spectrogram of the signal by comparing it to the energy computed using the time-domain data. We find that the total energy measured using the spectrogram is within $20\%$ of the energy measured using the time domain data. We then use the frequency evolution of the $f-$mode measured via the short-time Fourier transform to extract the energy from the time-frequency blocks associated with the $f-$mode oscillations using the spectrogram.  We find that the $f-$mode energies can be as high as $40\%$ of the total energy emitted in gravitational radiation during a core-collapse.  

We find that the energy associated with the $f-$mode oscillations typically increases with the progenitor mass. The energy also depends on the delayed explosion times and the success of explosion. Simulations having higher shock stall times before the onset of explosion emit more gravitational-wave radiation since the oscillations are excited for a longer time. Additionally, the energy of the $f-$mode also increases monotonically with the rotation rate of the core, up to a certain value of core rotation rate. Centrifugal forces dominate for faster core rotations, and cease the activation of the oscillations of the proto-neutron star. 

To understand how the detector noise will affect this analysis, we inject the gravitational-wave signals into simulated Cosmic Explorer and Einstein telescope noise and then extract the $f-$mode frequencies and measure the energies. We vary the distance to the source, but limit it to within the Milky Way galaxy. We find that for waveforms from three-dimensional simulations, we can measure the $f-$mode frequencies for sources up to $20$ kpc within an RMS error of 5 Hz, and the $f-$mode energies within $20 \%$ fractional error, when the gravitational-wave signal is assumed to be detected by a third-generation observatory. For waveforms from two-dimensional simulations with core rotation, we can measure the frequencies for sources up to $20$ kpc to within $2$ Hz RMS error and energies to within $10 \%$ fractional error.
 
Measurement of the frequencies and energies of the $f-$mode oscillations can provide us more information about the mechanism of the supernova explosion. We can also infer the central density of the proto-neutron star and the turbulence energy within the system.  

\section*{Acknowledgement}
We thank David Radice and Viktoriya Morozova for helpful discussions. AB acknowledges support from the U.~S.\ Department of Energy Office of Science and the Office of Advanced Scientific Computing Research via the Scientific Discovery through Advanced Computing (SciDAC4) program and Grant DE-SC0018297 (subaward 00009650), support from the U.~S.\ National Science Foundation (NSF) under Grants AST-1714267 and PHY-1804048 (the latter via the Max-Planck/Princeton Center (MPPC) for Plasma Physics), and support from NASA under award JWST-GO-01947.011-A. DV acknowledges support from the NASA Hubble Fellowship Program  grant HST-HF2-51520. A generous award of computer time was provided by the INCITE program, using resources of the Argonne Leadership Computing Facility, a DOE Office of Science User Facility supported under Contract DE-AC02-06CH11357. Also, some simulations were performed on Blue Waters under the sustained-petascale computing project, which was supported by the National Science Foundation (awards OCI-0725070 and ACI-1238993) and the state of Illinois. Blue Waters was a joint effort of the University of Illinois at Urbana--Champaign and its National Center for Supercomputing Applications. Finally, AB acknowledges computational resources provided by the high-performance computer center at Princeton University, which is jointly supported by the Princeton Institute for Computational Science and Engineering (PICSciE) and the Princeton University Office of Information Technology, and an allocation at the National Energy Research Scientific Computing Center (NERSC), which is supported by the Office of Science of the U.~S.\ Department of Energy under contract DE-AC03-76SF00098. SKK and ERC acknowledge support from the National Science Foundation through grant AST2006684, and ERC acknowledges additional support from the Oakridge Associated Universities through a Ralph E. Powe Junior Faculty Enhancement Award.

\bibliography{main}

\end{document}